\documentclass{article}

\usepackage{arxiv}

\usepackage[utf8]{inputenc} 
\usepackage[T1]{fontenc}    
\usepackage{hyperref}       
\usepackage{url}            
\usepackage{booktabs}       
\usepackage{amsfonts}       
\usepackage{nicefrac}       
\usepackage{microtype}      
\usepackage{lipsum}
\usepackage{graphicx}
\usepackage[version=4]{mhchem}
\graphicspath{ {./images/} }
\usepackage{multirow}
\usepackage{tabularx}

\title{One-dimensional self-organization of water molecules in proton conducting Andersson-Wadsley titanates.}

\author{Mathilde Arnaud$^{1\#}$, Guillaume Benas$^{1,2\#*}$, Narimane Meziani$^{3\#}$, Armel Descamps-Mandine$^{4}$,  \AND Claudie Josse$^{4}$, Sophia Akkari$^{8}$, Thierry Douillard$^{7}$, Sofia de Sousa Coutinho$^{1}$,  Mélanie De Vos$^{1}$, \AND Stéphane Holé$^{1}$, Gwenaëlle Rousse$^{5}$, Rémi Federicci$^{8}$, Sylvain Franger$^{3}$, \AND Paola Giura$^{6}$, Fabio Finocchi$^{2}$ and Brigitte Leridon$^{1,8*}$ \vspace{0.5cm}\\ $^\#$ \textit{These authors contributed equally.}\\ $^*$Corresponding authors : brigitte.leridon@espci.fr ; guillaume.benas@insp.jussieu.fr\vspace{0.5cm}\\ 
\small1. Laboratoire de Physique et d'Etude des Matériaux (LPEM) - Ecole Supérieure de Physique et Chimie Industrielles (ESPCI)\\ 
\small Université Paris Science et Lettres - Sorbonne Universités - Centre National de la Recherche Scientifique (CNRS)\\
\small2. Institut des Nanosciences de Paris (INSP) - Sorbonne Universités - Centre National de la Recherche Scientifique (CNRS)\\
 \small3. Institut de Chimie Moléculaire et des Matériaux d'Orsay (ICMMO) \\ \small Université Paris-Saclay - Centre National de la Recherche Scientifique (CNRS)\\
 \small4. Centre de Microcaractérisation Raimond Castaing \\ \small Université de Toulouse - INSA Toulouse - Centre National de la Recherche Scientifique (CNRS) \\
 \small5. Chimie du solide et énergie (CSE) - Collège de France \\
 \small6. Institut de Minéralogie, de Physique des Matériaux et de Cosmochimie (IMPMC)\\ \small Sorbonne Université - Centre National de la Recherche Scientifique (CNRS) - Muséum National d’Histoire Naturelle (MNHN)\\
 \small7. INSA Lyon - Université Claude Bernard Lyon 1 - MATEIS  - Centre National de la Recherche Scientifique (CNRS)\\
 \small8. Pioniq Technologies
}

\begin{document}
\maketitle

\pagestyle{fancy}


\keywords{Superionic conductors, solid ion conductors, confined water, fast proton conductors, self-organized water, perovskites}


\begin{abstract}

Layered alkali titanates with $\mbox{M}_2\mbox{Ti}_2\mbox{O}_5$ chemical formula (MTO, M=K,Rb) belonging to the Andersson-Wadsley perovskite family spontaneously incorporate water to form MTO.(H$_2$O)$_x$ compounds, which exhibit superionic conductivity. At very low hydration $x$, scanning electron microscopy evidences one-dimensional heterogeneous patterns oriented along $\vec{b}$ that are arranged in an orderly manner. At higher hydration, the material is observed to spontaneously exfoliate by creating (001) surfaces. 

Simulations carried out using Density Functional Theory reveal an ordered arrangement of the guest water molecules in Rb$_2$Ti$_2$O$_5$, with strong hydrogen bonds between the water molecules and the apical oxygen of the host crystal. At low hydration $x$, the water molecules form self-organized one dimensional (1D) double chains along $\vec{b}$. Further increase of the water content leads to the creation of hydrated (001)-surfaces that are made of densely packed water chains in agreement with the infrared spectroscopy measurements.  

Rb$_2$Ti$_2$O$_5$ exhibits highly anisotropic proton conductivity, with respect to the crystal orientation, with super-ionic conductivity along $\vec{b}$ reaching 3\,mS/cm at room temperature after hydration.

The combined observations and simulations suggest that these water chains are thus at the root of fast proton conduction, which is likely powered by a Grotthuss-like mechanism.  

\end{abstract}


\newpage
\section{Introduction}

In the search for sustainable electrical energy storage solutions, an important avenue of investigation lies in the discovery of innovative proton-conducting solid materials.
They can be used in particular for electrolysis and as proton-exchange membrane fuel cells (PEMFC).  
Indeed, although Nafion\textsuperscript{\textregistered} membranes are attractive for PEMFC, their high cost and limited temperature operation (below 80°C) make them irrelevant for large-scale applications.  
In order to bypass those limitations, the recent research is searching for new systems such as porous materials \cite{Meng_CSR_2017} or oxide proton conductors beyond perovskites \cite{FOP2021,Xu_2021,HUANG_2023}, as an alternative to Li$^+$-based conductors.

Protons, on the contrary, owing to their small ionic radius engender ultra-fast diffusion kinetics, with a mobility of about $3\times10^{-3}$cm$^2$/V/s, leading to high power density capabilities \cite{Xu_2021}. 
In addition, the Hydrogen ability to form both ionic, covalent and hydrogen bonds with several electrode materials enables fast-charging and discharging processes. All-solid proton batteries would combine the advantages of proton conduction to those of all-solid design: absence of potentially leaking elements, stability, reliability, larger working potential and high degree of integrability. 
This would lead to improved energy density and enable designing on-chip micro-batteries. 

For all these devices, solid-state proton conductors with very high conductivity are required. Perovskite-derived materials that incorporate and dissociate water have been extensively studied \cite{FOP2021}, but, to date, they display conductivity of the order of at most 10$^{-3}$~S/cm at 200°C (See Figure ~2 in \cite{FOP2021}). 
Crystals that can incorporate judiciously organized water molecules, so to conduct protons through Grotthuss mechanism \cite{Grotthuss_1806,AGMON_1995}, also constitute promising candidates.
They also show a wider electrochemical and temperature stability window with respect to liquid aqueous electrolytes. 

In this respect, porous materials where guest water molecules incorporate as chains that could provide fast proton conduction -- such as metal-organic frameworks (MOF) -- have been considered but suffer from relatively low stability \cite{Meng_CSR_2017,Shimizu_2013,Zhao_CM_2014}. Covalent organic framework (COF) are found to be more stable proton conductors \cite{Chandra_JACS_2014}.  
Zeolites have also been investigated as proton conductors, in particular for fuel cell applications \cite{YEUNG_CT_2024}.  
As an alternative, two-dimensional lamellar inorganic materials are extremely promising as super-ionic conductors at the nanometer scale \cite{Sun_2018}.

Here we investigate alkali titanates from the Andersson-Wadsley family \cite{andersson_five_1960}, systems that are found to behave with marked differences from those that have been studied so far.
More precisely, we describe and analyze the interplay between alkaline titanium oxides with chemical formula $\mbox{M}_2\mbox{Ti}_2\mbox{O}_5$ (MTO), with M=(Rb, K) and guest water molecules. 
Indeed these materials exhibit several desirable characteristics: they show a marked anisotropy with some properties that are comparable to those of two-dimensional systems; they have a high water affinity even in the absence of extrinsic defects; and exposure to water vapor triggers superionic conduction properties.
 
By combining electron microscopy, first-principle simulations and infra-red spectroscopy, we show that water molecules are absorbed by the layered structure of the perovskite-derived $\mbox{Rb}_2\mbox{Ti}_2\mbox{O}_5$  (RTO) or $\mbox{K}_2\mbox{Ti}_2\mbox{O}_5$  (KTO) compounds and induce fast proton conduction. 
The measured ionic conductivity of these layered oxides is found to be strongly anisotropic, and amounts to $0.27-3~\mbox{ mS/cm}$ at room temperature along $\vec{b}$. 
Therefore, they represent extremely promising building blocks for solid-state energy storage devices.

\section{Results and discussion}

\subsection{MTO superionic conductors}

The MTO family, which is part of the Andersson-Wadsley family, was discovered in the early 60s when it was synthesized and its structure determined. 
\cite{andersson_five_1960}
Its structure is rather peculiar due to the relatively unusual five-fold Ti coordination number with O. 
However, the electric properties of these compounds were not explored at that time. 
Complex admittance characterization of $\mbox{M}_2\mbox{Ti}_2\mbox{O}_5$.(H$_2$O)$_x$  crystals was indeed performed only recently and very high values of the equivalent relative permittivity (up to $10^9$) were observed and interpreted as resulting from a colossal accumulation of charge on the surface due to super-ionic conduction\cite{Federicci_PRM_2017,federicci_memory_2018,DESOUSACOUTINHO201972, RANI2020126784}. It was then demonstrated that the ionic conduction is due to the hygroscopic character of the material that spontaneously incorporates water from the atmosphere, which presumably triggers proton conduction\cite{desousacoutinho_SSI_2021}.

A recent study performed on powders elucidates that hydration in presence of CO$_2$ leads to partial carbonation of the compounds by creation of MHCO$_3$ and then M$_2$CO$_3$, which is only reversible at very high temperatures (800°C)\cite{Meziani_2024}.

\subsection{\label{micro} Electron microscopy}

\begin{figure}[t]

\parbox[t]{0.19\textwidth}{
    {\large\bfseries A}\\[1mm]
    \includegraphics[scale=0.212]{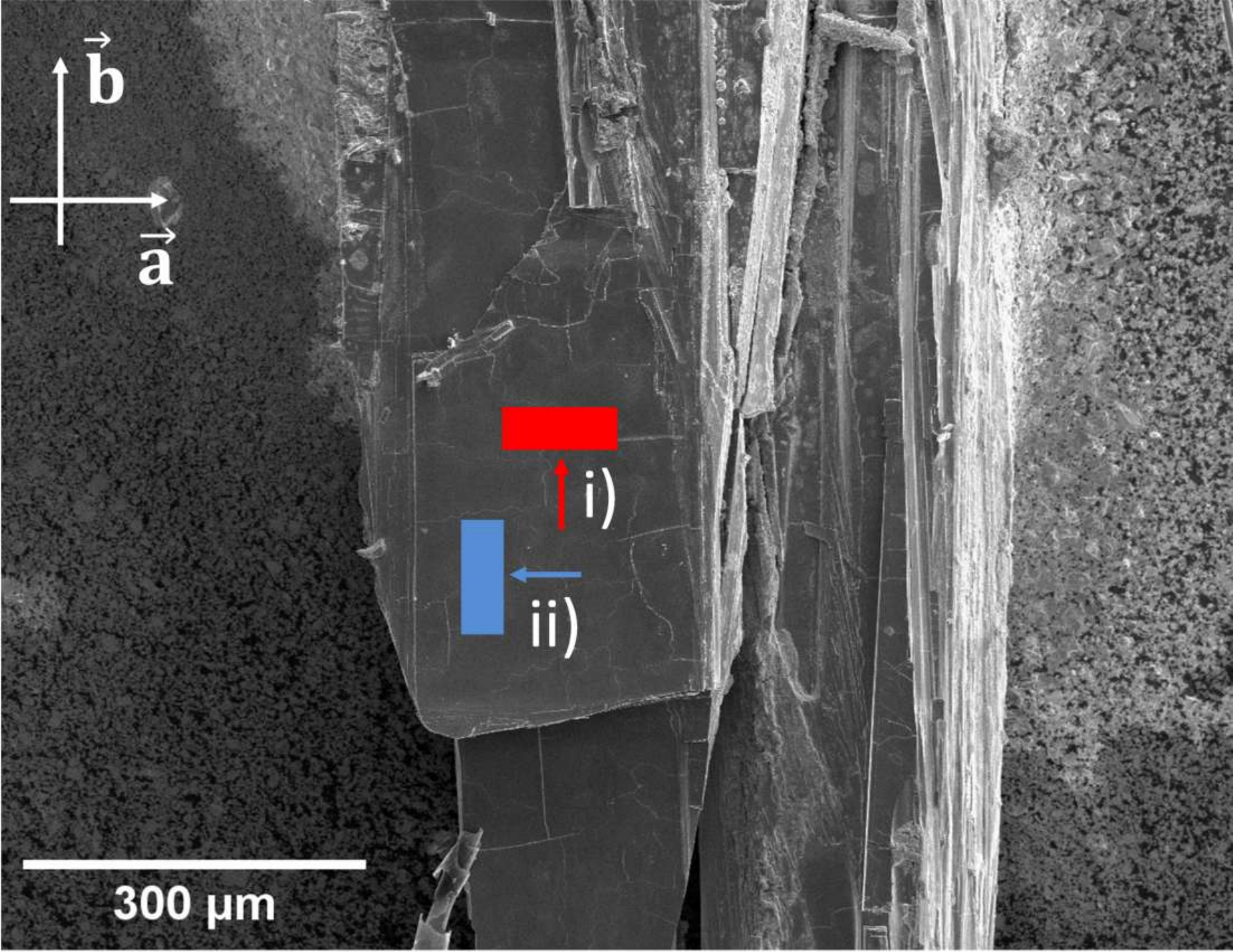}
}\hspace{2.5 cm}
\parbox[t]{0.19\textwidth}{
    {\large\bfseries B}\\[1mm]
    \includegraphics[scale=0.225]{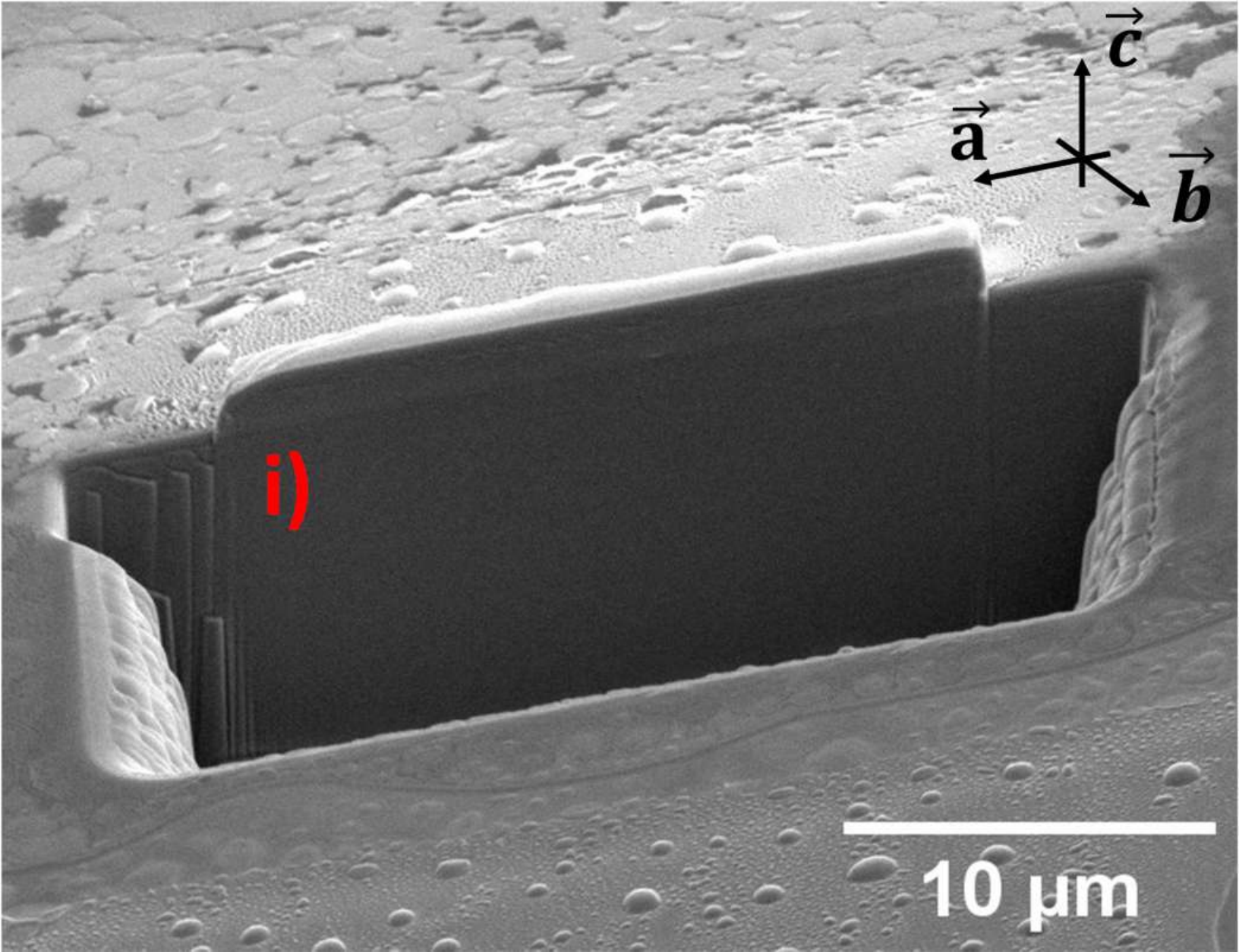}
}\hspace{2.5 cm}
\parbox[t]{0.19\textwidth}{
    {\large\bfseries C}\\[1mm]
    \includegraphics[scale=0.225]{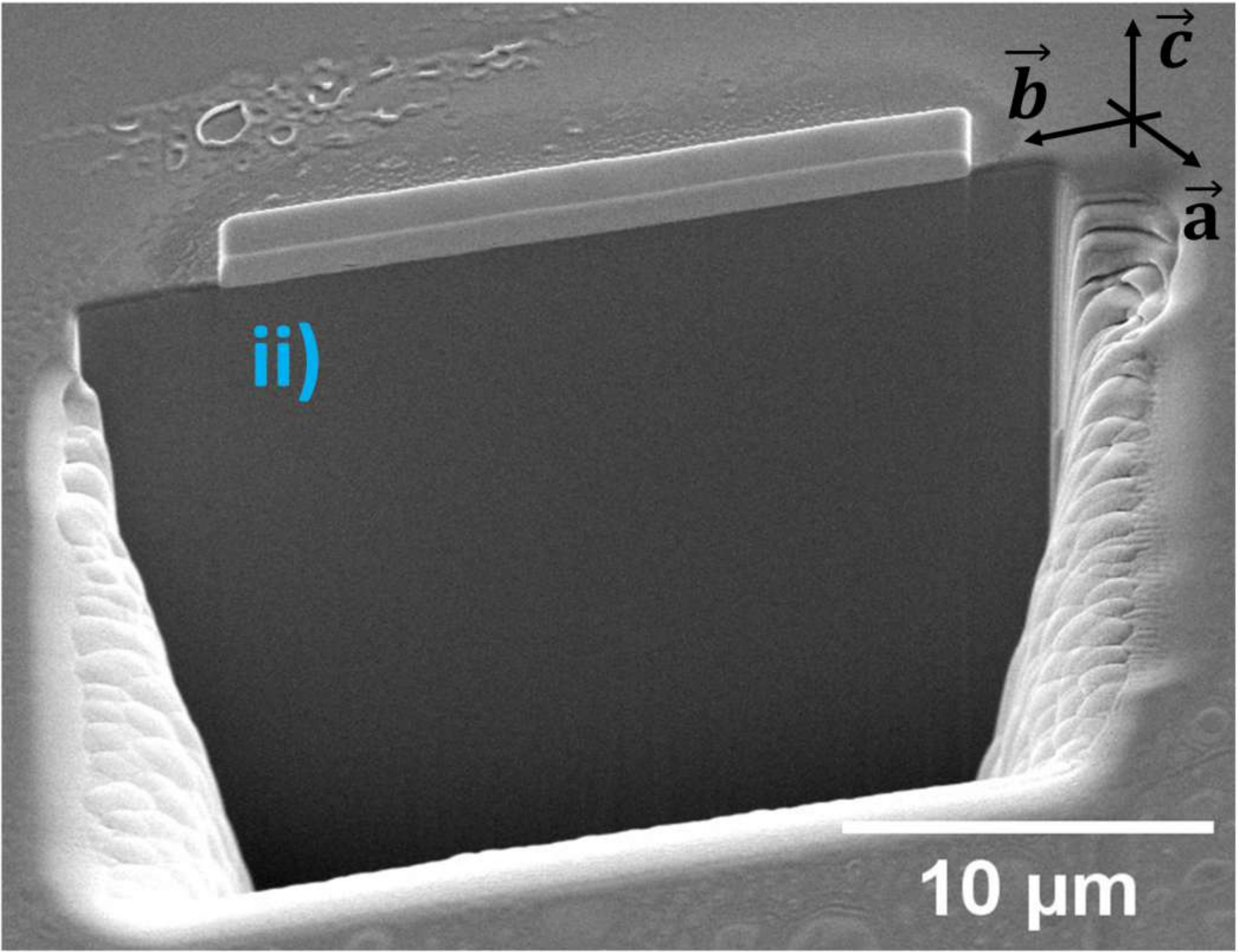}
}\\
\vspace{0.25 cm}
\hspace{5.5 cm}
\parbox[t]{0.19\textwidth}{
    {\large\bfseries D}\\[1mm]
    \includegraphics[scale=0.225]{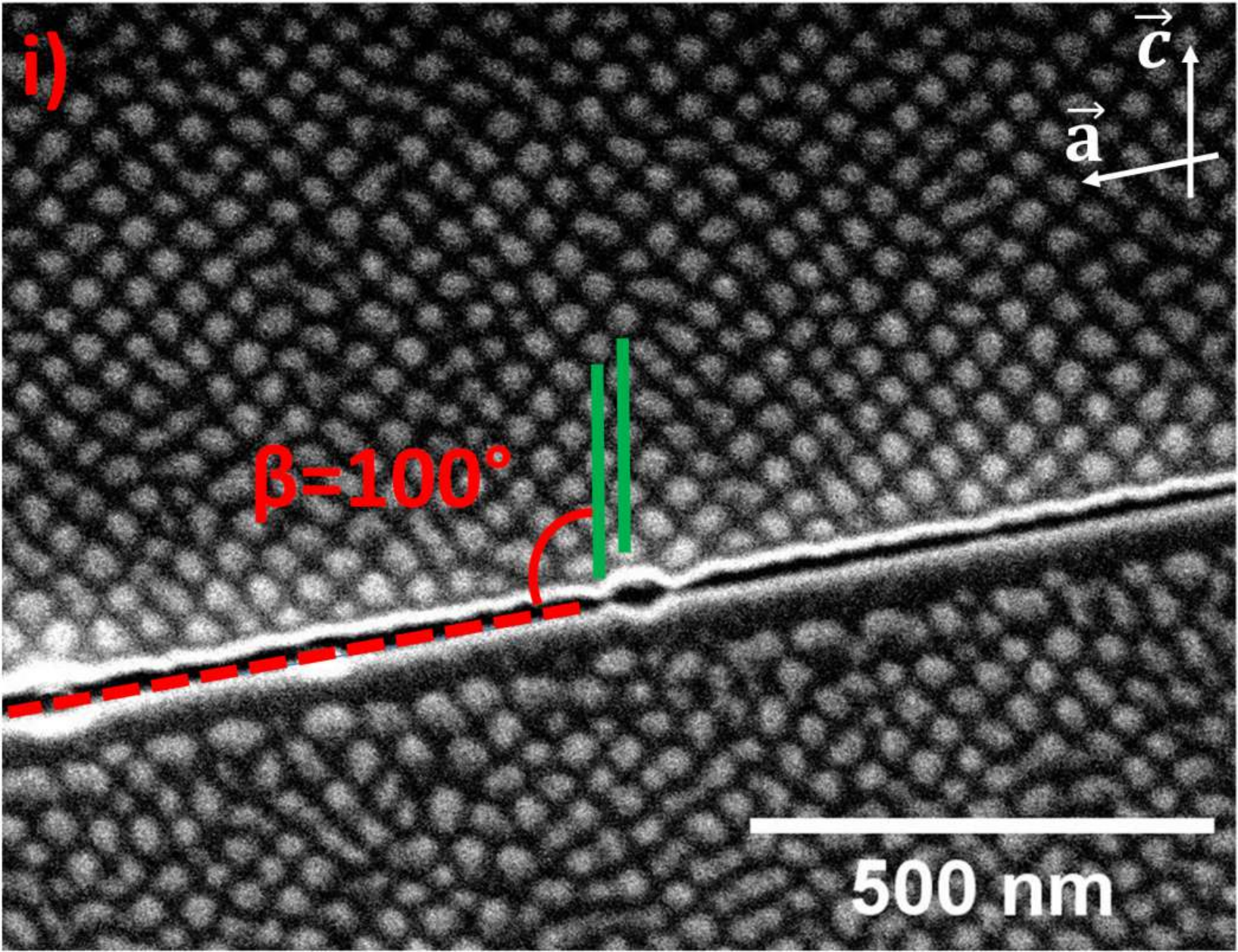}
} \hspace{2.4 cm}
\parbox[t]{0.19\textwidth}{
    {\large\bfseries E}\\[1mm]
    \includegraphics[scale=0.225]{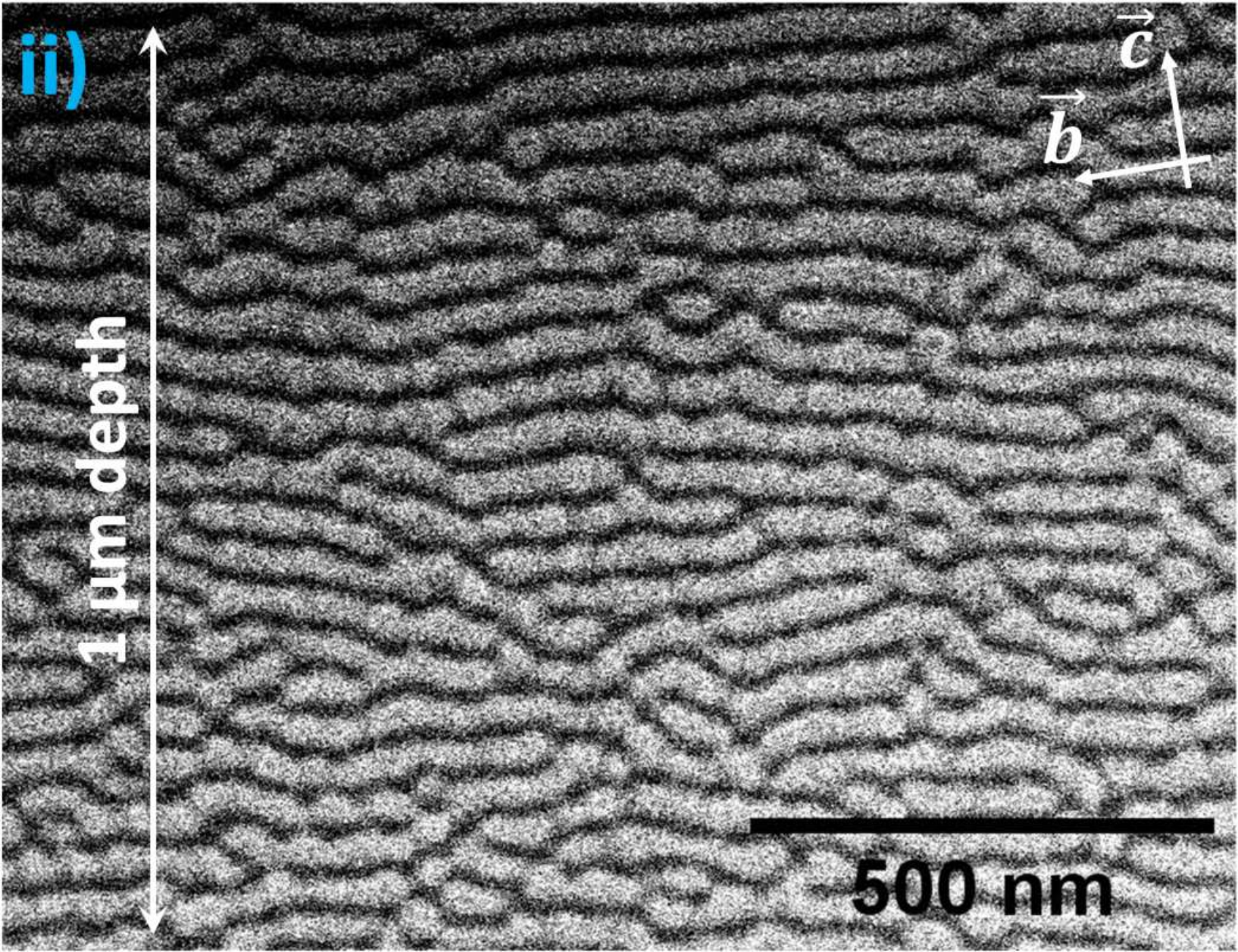}
}
\caption{\label{fig1MEB} A) SEM image of an RTO single crystal hydrated for less than 9\,min in the lab atmosphere. Two orientations were selected for the FIB-milled platelets: one perpendicular to $\vec{b}$ represented by the red frame (ii) RTO1b), and the other perpendicular to $\vec{a}$ represented by the blue frame (ii) RTO1a). B) and C) SEM images of each oriented FIB-milled platelets (RTO1b and RTO1a). D) SEM image of an enlargement of RTO1b showing a highly ordered pattern of white dots. The crystallographic orientation was deduced from the crack (red dash line) which corresponds to the (001) cleaving plane. The green vertical lines represent the crystallographic orientation $\vec{c}$. E) SEM image of an enlargement of RTO1a showing elongated horizontal white zones. The crystallographic orientation was deduced from the surface of the platelet. An classical SEM Everhart-Thornley detector (ETD) was used for images A), B) and C) and a Through the Lens Detector (TLD) for images D) and E). All images were obtained in Secondary Electron (SE) mode.}
\end{figure}

Both RTO and KTO samples were analyzed by electron microscopy. 
Focused ion beam (FIB) milling was performed inside a FIB-equipped scanning electron microscope (FIB-SEM) to obtain oriented platelets from single crystals with various degrees of hydration and carbonation (see Figure \ref{fig9MEB} in Methods Section). Figure \ref{fig1MEB}A shows a RTO single crystal exposed for less than 9\,min to the lab atmosphere (of about 45\% humidity) before being introduced into the FIB chamber. This crystal can be considered as "lightly hydrated" due to the short exposure time. Two different platelets of RTO were milled in two different directions: one perpendicular to $\vec{b}$ (RTO1b, red frame in Figure \ref{fig1MEB}A) and the other perpendicular to $\vec{a}$ (RTO1a, blue frame in Figure \ref{fig1MEB}A). For KTO single crystals only platelets oriented perpendicular to $\vec{b}$ were prepared (KTO1b and KTO2b). The obtained platelets were then analyzed by scanning electron microscopy (SEM) in Secondary Electron (SE) mode.

\paragraph{Rb$_2$Ti$_2$O$_5$ \label{sec:RTO}}

The lightly hydrated RTO sample shown in Figure \ref{fig1MEB} was studied under two different orientations. A zoomed SEM image of the (010) surface of RTO1b is shown in Figure \ref{fig1MEB}D, corresponding to the plane represented in Figure \ref{fig1MEB}B. RTO1b shows bulk heterogeneities extending over more then 1 micron in depth marked by light and dark zones, arranged in an orderly manner. 
A crack is also evident about the center of Figure \ref{fig1MEB}D, which corresponds to a (001) crystal cleavage plane. 
The same types of contrasts (white dots) was observed on several other RTO samples on surfaces oriented perpendicular to $\vec{b}$ and appeared to have an almost square arrangement. 
The shortest center-to-center distance between the white dots in the network amounts to  45 $\pm$ 5 nm and is quite uniform within the material. In order to further characterize this periodic network, the Fast Fourier Transform (FFT) of one image was calculated and is presented in Figure S1 of the Supporting Information. This confirms the existence of a long-range ordering with an overall square symmetry for the white dots, with angles around 90°$\pm$ 3°. 
It is noteworthy that in Figure \ref{fig1MEB}D one of the diagonals of the nearly-square pattern forms with the crystallographic direction $\vec{a}$ of RTO (as materialized by the visible cleavage plane) an angle of about 100°, which corresponds to the $\beta$ angle of the monoclinic crystal \cite{Federicci_AC_2017}. Therefore, we identify the diagonal of the white dot contrasts with the crystal $\vec{c}$ axis. 
This underlies that the orientation of the white dots is correlated with the crystallographic structure of RTO.

The SEM image of the RTO1a platelet is presented in Figure \ref{fig1MEB}E. Here, on the (100) surface, the bulk heterogeneities of light and dark zones are also observed, although it appears they appear very different from those seen on the (010) surface (Figure \ref{fig1MEB}D). Here indeed the contrast pattern is given by the alternation of elongated almost parallel white and dark zones, roughly oriented along $\vec{b}$, with typical lengths of the order of 900\,nm. The distance center-to-center of the white elongated objects along $\vec{c}$ is of the order of 45\,nm $\pm$ 5\,nm, which corresponds to the nearest-neighbor distance seen in the RTO1b between two white dots.

In addition, a separate study was performed on another hydrated RTO single crystal (denoted RTO2b) under low-energy SEM imaging (see Figure S2 left panel in the Supporting Information). This RTO2b sample manifests an hexagonal white dots arrangement for a surface perpendicular to $\vec{b}$. Also, both STEM-HAADF and HRTEM images (see Figure S2 in the Supporting Information) were carried out on RTO2b. The HRTEM image demonstrated a fully preserved atomic network consistent with the RTO structure (see Figure S2 center panel) whereas the STEM-HAADF image confirmed that the white dot pattern exhibited local hexagonal symmetry (see Figure S3 in the Supporting Information).  

\paragraph{K$_2$Ti$_2$O$_5$ \label{sec:KTO}}

\begin{figure}
\begin{center}
 \includegraphics[scale=0.44]{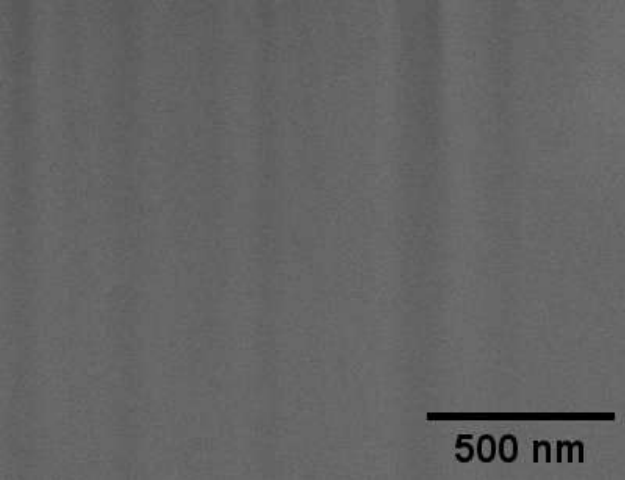}
\hspace{1cm}
\includegraphics[scale=0.22]{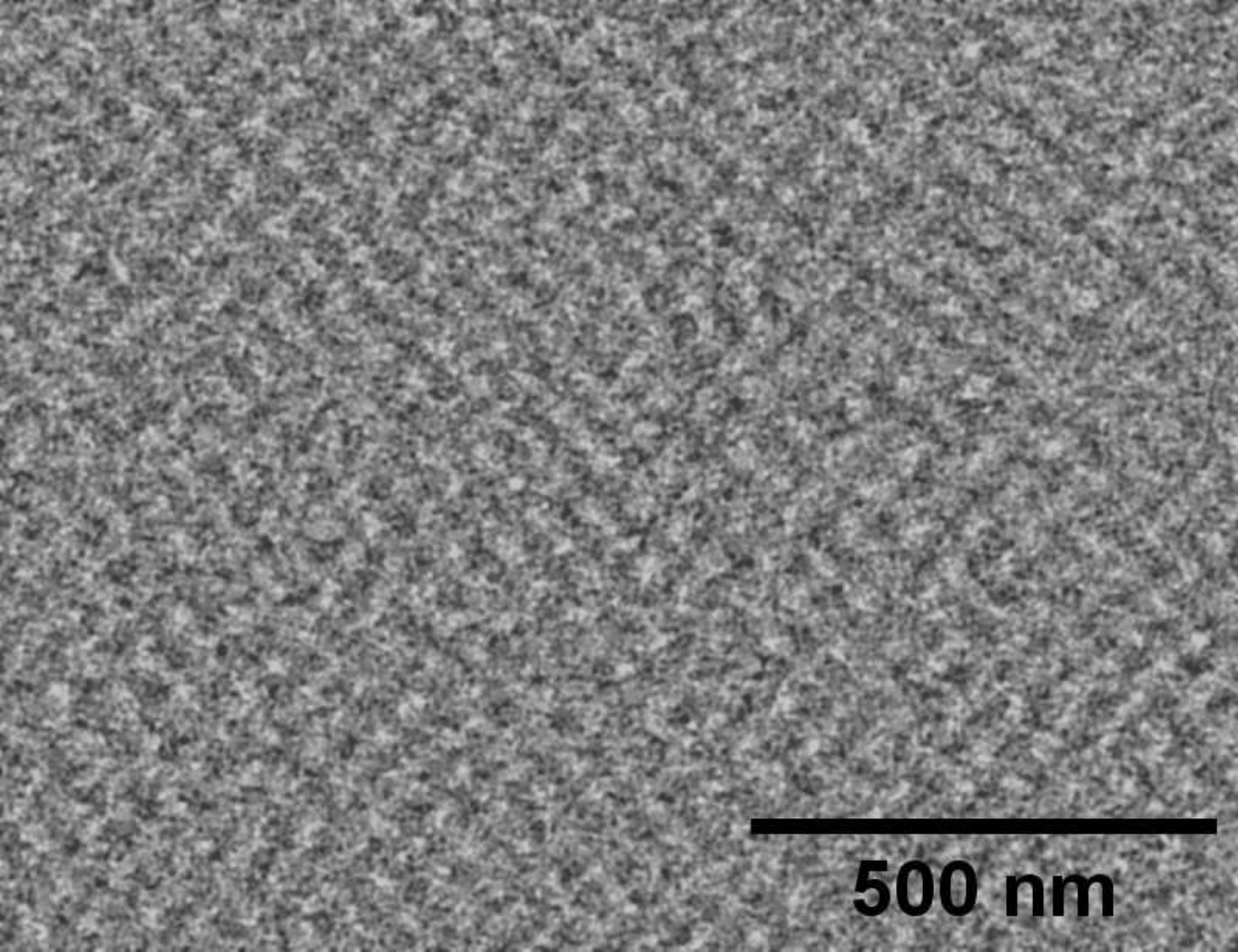}
\caption{\label{figKTOMEB} Left: SEM image of surface perpendicular to $\vec{b}$ of a KTO single crystal exposed less than 9\,min to the laboratory atmosphere (KTO1b). No heterogeneous volume composed of light and dark contrasts is observed for this sample. Right: SEM image of KTO exposed for 1h30 at the laboratory atmosphere (about 45\% of humidity) and then placed into the FIB-SEM chamber (KTO2b). A heterogeneous volume composed of light and dark contrasts is observed for this sample. Both SEM images were acquired using a TLD detector in Secondary Electron (SE) mode.}   
\end{center}
\end{figure}

A similar study was conducted on a KTO single crystal similarly exposed to the lab atmosphere for less than 9\,min (KTO1b, left panel of Figure \ref{figKTOMEB}) and for 1h30 (KTO2b, right panel of Figure \ref{figKTOMEB}). Differently to RTO1b, no contrast heterogeneities are observed on the (010) surface in the case of KTO1b, while they appear in the KTO2b, thus underlying a similar overall behavior caused by air exposure on different time
scales. In Figure \ref{figKTOMEB} right panel, the spurious additional noise observed is caused by the water cooling system of the sample holder which generated instabilities during the SEM imaging. 

Exposition to ambient atmosphere is known to trigger hydration and carbonation with very different hydrating rates for RTO and KTO. DRX measurements \cite{Meziani_2024} performed on KTO and RTO powders show that the former is about 10 time slower in reacting with water and CO$_2$ than the latter. Taking this hydration rate into account, the 1h30 exposure time for KTO2b is roughly equivalent to a 9 min exposure time for RTO, which corresponds to the exposure time of RTO1b (the RTO sample with the shortest exposure time). Therefore it is not surprising that the heterogeneous contrasts while absent in KTO1b are present in RTO1b.

In summary, electron microscopy studies of the (010) and (100) surfaces of RTO and (010) surface of KTO with different water contents revealed the presence of anisotropic heterogeneous contrasts, which appear to originate from hydration. These contrasts form a highly ordered pattern of white dots on the (010) surface, which extend into one-dimensional contrasts along $\vec{b}$.

\subsection{\label{DFT} First-principles simulations.}

In order to understand how water molecules incorporate in Rb$_2$Ti$_2$O$_5$, we carried out first-principles simulations within the Density Functional Theory (DFT). The computational details are given in the method section.

\begin{figure}[h!]
\begin{center}
\includegraphics[width=1\textwidth]{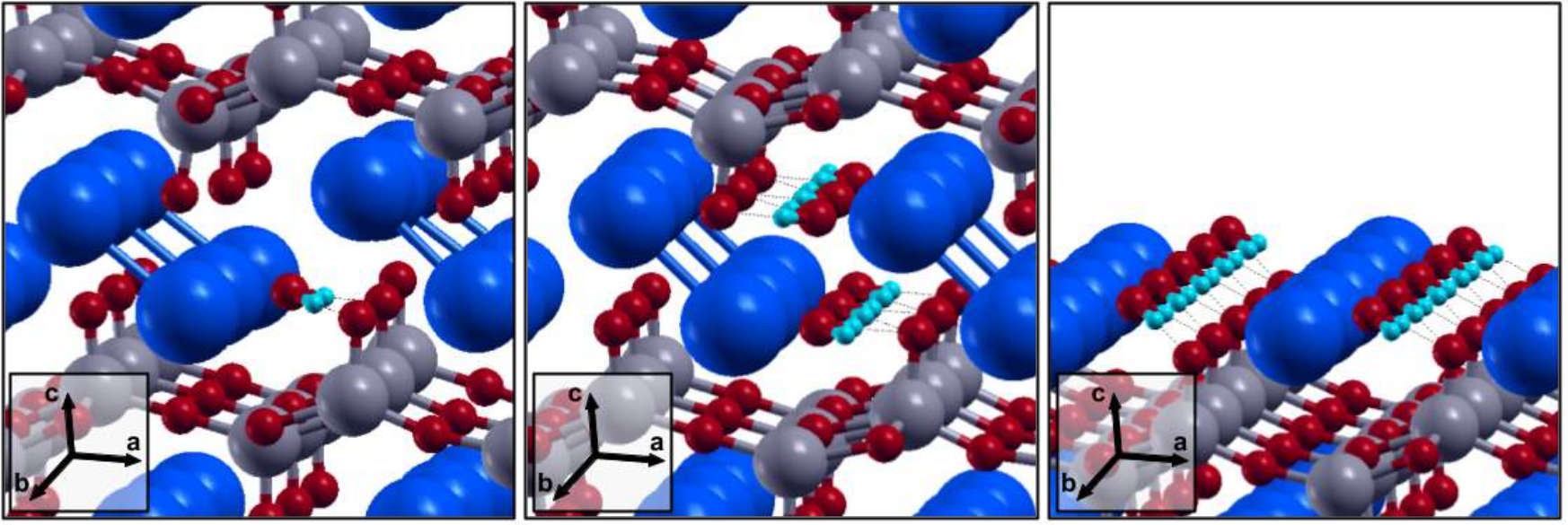}
\caption{\label{fig:hyd.img} \underline{\textbf{Configurations of hydrated RTO}} illustrated using XCrysDen \cite{xcrysden}. 
Left panel: the most stable adsorption site, RbO-coplanar configuration (v1). Middle panel: arrangement of the water molecules in two facing RbO-coplanar chains as described in the text. Right panel:  (001) surface uniformly covered with water molecules. The Ti atoms are represented in gray, the Rb atoms in dark blue, the O atoms in red and the H of the water molecules in light blue.}
\end{center}
\end{figure}

For an isolated water molecule, after an extensive search we identify two distinct bulk adsorption sites. In the most stable configuration O$^{(w)}$ binds to Rb atoms, and the two protons form short and strong H bonds with two apical O$^{(1)}$ that are aligned along $\vec{b}$ (see Figure \ref{fig:hyd.img}). The molecule lies in a plane which is nearly normal to $\vec{c}$ with its molecular axis pointing in the direction along $\vec{a}$, and is therefore denoted as "RbO-coplanar". The molecule can also be adsorbed in a metastable state by forming H bonds with two apical O$^{(1)}$ within facing RbO planes, denoted as "bridging" (see Figure S4 in the Supporting Information). Detailed properties of the two configurations are listed in the Table S3 of the Supporting Information.
In both configurations, the short and strong hydrogen bonds stretch the covalent OH bonds within the molecule, resulting in a red-shift of the  $\mathrm{O}^{(w)}-\mathrm{H}$ stretching modes, which vibrate around 3000~cm$^{-1}$ and display a characteristic doublet (Table \ref{tab:hyd_freq}). Notably, in the RbO-coplanar configuration, the molecule exhibits an inversion in the frequency ordering of the symmetric and antisymmetric stretching modes relative to the gas phase, resulting from interactions with the crystal’s neighboring atoms. In contrast, in the bridging configuration, no symmetric stretching mode is observed due to asymmetric hydrogen bonding (see the Supporting Information for details).

\begin{table}[h!]
\centering
\footnotesize
\setlength{\tabcolsep}{4.2pt}
\begin{tabular}{|c|c||c|c||c|c||c|c|}
\hline
\multicolumn{2}{|c||}{\textbf{Configuration}} &
$\mathbf{\nu_1}$ &
$\max_{\alpha}\{\hat{\mathbf{p}}_{\alpha}^{\nu_{1}}\}$ &
$\mathbf{\nu_{2}}$ &
$\max_{\alpha}\{\hat{\mathbf{p}}_{\alpha}^{\nu_{2}}\}$ &
$\mathbf{\nu_{3}}$ &
$\max_{\alpha}\{\hat{\mathbf{p}}_{\alpha}^{\nu_{3}}\}$ \\
\hline\hline

\textbf{Gas} &
Single molecule &
1591 &
/ &
3650 (SS) &
/ &
3765 (AS) &
/ \\
\hline\hline

\multirow{3}{*}{\textbf{Bulk}}
& Single molecule, bridging
& 1505
& $\vec{b}$
& 2784
& $\vec{c}$
& 2925
& $\vec{a}\sim\vec{b}$ \\
\cline{2-8}

& Single molecule, RbO-coplanar
& 1658
& $\vec{a}$
& 2777 (AS)
& $\vec{b}$
& 2879 (SS)
& $\vec{a}$ \\
\cline{2-8}

& Double chains, RbO-coplanar
& 1699
& $\vec{a}$
& 2980 (AS)
& $\vec{b}$
& 3112 (SS)
& $\vec{a}$ \\
\hline\hline

\multirow{2}{*}{\textbf{(001) surface}}
& Single molecule, RbO-coplanar
& 1688
& $\vec{a}$
& 2987 (AS)
& $\vec{b}$
& 3049 (SS)
& $\vec{a}$ \\
\cline{2-8}

& Chains, RbO-coplanar
& 1729
& $\vec{a}$
& 2982 (AS)
& $\vec{b}$
& 3113 (SS)
& $\vec{a}$ \\
\hline
\end{tabular}

\caption{\label{tab:hyd_freq}
Water molecule vibrational frequencies (cm$^{-1}$): bending mode $\mathbf{\nu_1}$ and stretching modes $\mathbf{\nu_{2}}$ and $\mathbf{\nu_{3}}$, with stretching character indicated when relevant (SS = symmetric stretching, AS = antisymmetric stretching). The direction of the largest component of polarization $\max_{\alpha}\{\hat{\mathbf{p}}_{\alpha}^{\nu}\}$ associated to each mode is expressed in terms of lattice vectors. These are computed for several configurations at the harmonic level (see Methods Section for computational details).}
\end{table}

At higher water concentrations, water molecules preferentially adsorb on neighbor RbO-coplanar sites, forming H$_2$O chains that align along $\vec{b}$ on facing RbO$^{(1)}$ planes, with antiparallel dipoles oriented almost parallel to $\vec{a}$. This one-dimensional ordering is thermodynamically favored over isolated or randomly distributed configurations, strongly stabilizing the system while minimizing structural constraints within the material (see Figure S5 in the Supporting Information). 
Water adsorption thus gives rise to a dense network of alternating $\mathrm{O}^{(w)}-\mathrm{H}$ and $\mathrm{O}^{(w)}-\mathrm{Rb}$ covalent bonds as well as tight hydrogen bonds with apical oxygen (see Figure S6 in the Supporting Information).
The resulting OH stretching frequencies ($\sim 3000$~cm$^{-1}$) are therefore strongly red-shifted  (see Table \ref{tab:hyd_freq}).

The computed negative adsorption enthalpy is consistent with the highly hygroscopic nature of the material; moreover, the self-organization of water in one-dimensional ordered chains \cite{Yoon2025} rather than in closed-loop or disordered networks, can be prone to proton conduction along the chain direction.

Notably, the insertion of molecules within the bulk induces significant mechanical stress, which could in principle be accommodated by an expansion of the inter-layer spacing. However, the absence of experimental evidence for changes in lattice parameters indicates an extremely low concentration of the water molecules within in the bulk. This is supported by the effective repulsive interactions as found between the double chains (Figure S7, Supporting Information), suggesting a preferential ordering of the double chains into regular array over long distances. Such an arrangement would minimize mechanical stress while preserving the structure of the crystal lattice and would correspond to an hydration content much lower than what is experimentally detectable.

For higher water content, another process could be at work to reduce the elastic energy within the material: the formation of hydrated (001) surfaces. 
Indeed, as the hydration rate increases, the hydrated (001) surface becomes more stable than the hydrated bulk, which is consistent with the observed exfoliation of the material under humid conditions \cite{Benas2025}. 
{Water molecules adsorb on the (001) surface in the same fashion as in the bulk, forming chains aligned along $\vec{b}$ (Figure \ref{fig:hyd.img}). 
Consequently, the OH stretching frequencies closely resemble those of the bulk double chains, in very good agreement with the experimental observations (see Table \ref{tab:hyd_freq}).
At constant volume, surface hydration is thermodynamically favored over bulk hydration.
The former leads to significantly lower internal stress within the material than the latter hydration mode (see in the Supporting Information Figure S8).  
At further surface hydration (1 Mono-Layer ML), water can adsorb atop the surface plane or in the layer beneath the surface. 
According to preliminary calculations, the latter configuration is more stable and results in a outward relaxation of the hydrated surface bi-layer along $\vec{c}$ from the dry bulk layers. 
In both configurations, the water chains on the (001) outermost surface plane is little affected, presumably because of the formation of strong bonds with the neighboring atoms \cite{Benas2025}.} 
In the case of high water coverage at the surface ($\ge$1 ML), the outermost Rb ions can also undergo a remarkable outward relaxation.
This might be the sign of the onset of Rb dissolution, which likely favors the formation of Rb(OH)$_n$ complex as well as the incorporation of protons into the structure, in substitutional sites (H$_\mathrm{Rb}$) or bound to apical O ions. 

We point out that the previously mentioned simulations only consider the interaction between Rb$_2$Ti$_2$O$_5$ and water, neglecting the presence of atmospheric CO$_2$ and the associated reaction products. 
As reported in \cite{Meziani_2024}, pure hydration is accomplished under CO$_2$-free atmosphere, in order to avoid the formation of carbonates. 
Indeed, given the alkaline nature of the (001) termination, the formation of carbonic acid (H$_2$CO$_3$), bicarbonate (HCO$_3^-$) or Rubidium bicarbonate (RbHCO$_3$) in the presence of H$_2$O and CO$_2$ is expected. Interestingly, the interaction of these species with the structure might promote proton mobility along the short H bonds that are present in carbonates \cite{Larvor,odin}. 
A detailed investigation of the hydration in the presence of carbonates will be addressed in future studies.

\subsection{\label{IR} Infra-red spectroscopy}

To probe the signatures of water molecules within the compound and compare them to the predictions from the DFT calculations, polarized infrared spectroscopy measurements were carried out on single crystals with different degrees of hydration.
\begin{figure}[h!]
\centering
\includegraphics[width=0.8\linewidth]{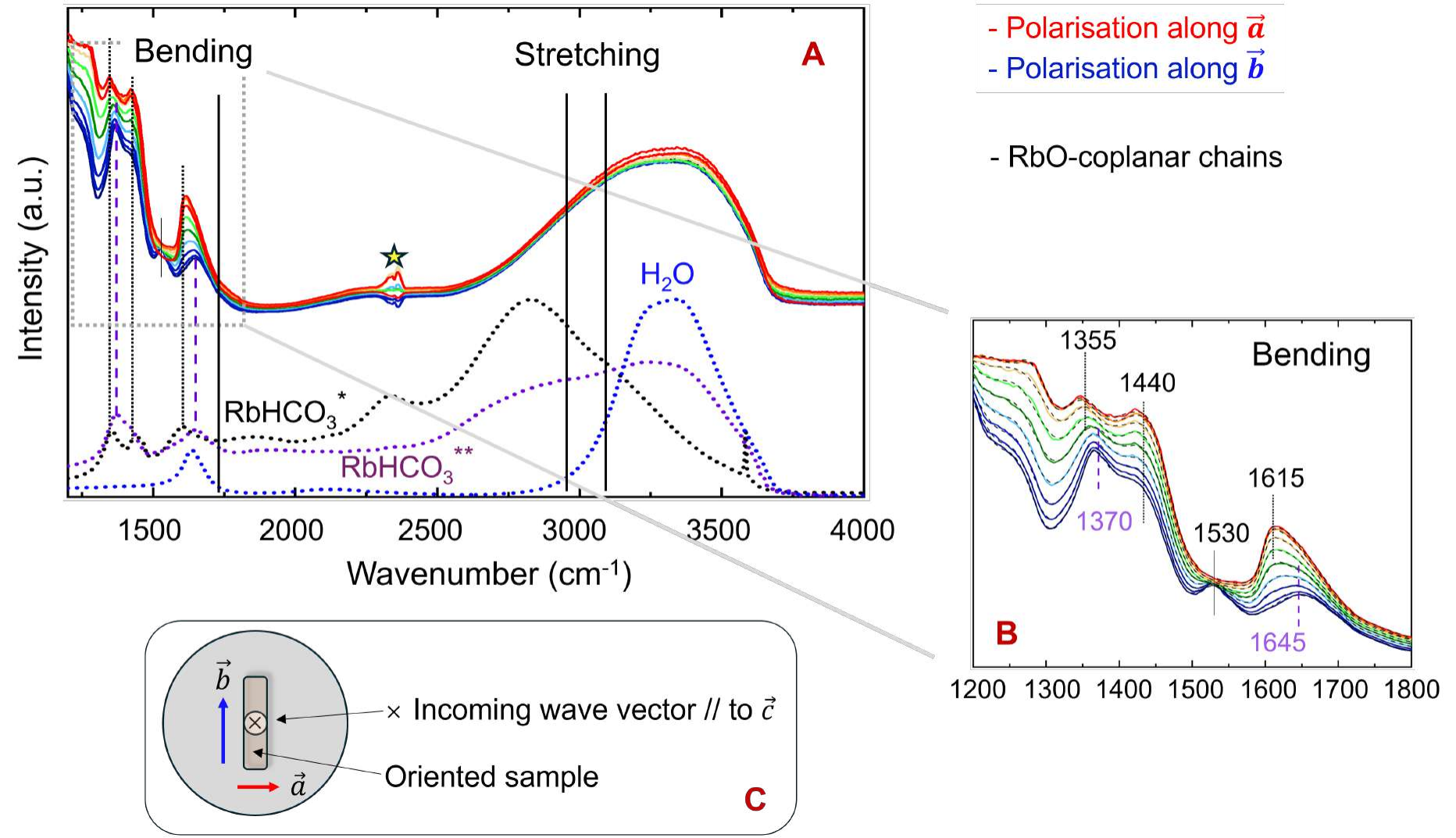}

\caption{
\label{FTIR} A) Infrared spectra of an RTO single crystal hydrated by exposure to ambient air for less than one hour. The RTO spectra were acquired as a function of polarization, ranging from $\vec{b}$ (blue solid curve) to $\vec{a}$ (red solid curve) in 10° increments. The star indicates the non-compensated absorption of the atmospheric CO$_2$ antisymmetric stretching vibration. The dotted curves correspond to the Attenuated Total Reflection (ATR) spectra of liquid water (blue), RbHCO${_3}^*$ (hydrated and carbonated rubidium dioxide powder, black) and RbHCO${_3}^{**}$ (hydrated solvated and carbonated rubidium dioxide, purple). The vertical solid lines indicate the positions of the DFT bending and stretching modes for the RbO-coplanar chains configuration (refer to Table \ref{tab:hyd_freq}). The vertical dashed and dotted lines correspond to the modes described in the text, the frequencies of which are reported in the inset B. Excerpt C shows the sample orientation and the experimental alignment geometry.}
\end{figure}

Figure \ref{FTIR} displays the infrared absorption spectra measured on the less hydrated RTO sample (solid curves) in the mid-infrared range where the bending and stretching vibration modes of water take place. 
The sample (0.54 mm thick, 3 mm long, and 0.6 mm wide), was aligned with $\vec{c}$ parallel to the incoming probe $\vec{k}$ vector thus to explore polarizations within the plane containing the $\vec{a}$ and $\vec{b}$ crystal axes (see Figure \ref{FTIR}C). Figure \ref{FTIR} also reports the Attenuated Total Reflection (ATR) absorption spectra of pure liquid water and carbonated rubidium hydroxide (RbHCO$_3$) at different degrees of hydration. 
The RTO single crystal as shown in Figure \ref{FTIR} was hydrated by exposure to the laboratory atmosphere for the time of the experiment (less than 1 hour). 
It is worth stressing here that polarized measurements have been repeated over a period not exceeding three hours, in order to check the reproducibility of the measurements and identify the transient features that evolve in time.

The infrared spectra in Figure \ref{FTIR} show a noticeable dependence on polarization, mainly in the 1000–2000 cm$^{-1}$ range, where H$_2$O bending modes and carbonate-related modes are expected. Specifically, the vibrational modes between 1300 and 1500 cm$^{-1}$ are related to adsorbed HCO${_3}^-$ groups with different degrees of hydration (black and purple dotted curves). A detailed description of the carbonate and water co-adsorption will be the subject of a forthcoming work; here it is worth noting that in the recorded spectra the mode at 1440 cm$^{-1}$ has maximum amplitude for polarized light $\parallel\vec{a}$, while that at 1370 cm$^{-1}$ has its maximum for polarized light $\parallel\vec{b}$, supporting a slanted alignment of the HCO${_3}^-$ groups with respect to (001) planes. 
In the 1500-1700 cm$^{-1}$ range (see inset B in Figure \ref{FTIR}) there are three more peaks at 1530 cm$^{-1}$, 1615 cm$^{-1}$ and 1645 cm$^{-1}$. 
The former evolves with time and has recently been reported by Münst and coworkers \cite{munst2021infrared}, which they attributed to the evolution of CO${_3}^{2-}$ group vibrations in the presence of water. 
Remarkably, the peaks at 1615 cm$^{-1}$ and 1645 cm$^{-1}$ display a reproducible behavior with polarization.
The former reaches its maximum intensity for polarization along $\vec{a}$ and is totally screened along $\vec{b}$, while the latter is hidden in the asymmetric shape of the 1615 cm$^{-1}$ peak along $\vec{a}$ and does not seem to be sensitive to polarization.
These features are the signatures of the water bending modes. 
The only water configuration that is consistent with the experimental results is the RbO-coplanar, for which the bending mode is polarized along $\vec{a}$ (contrary to the bridging configuration). 
In contrast, the larger structure at 1645 cm$^{-1}$ is insensitive to polarization and suggests the presence of non-oriented water molecules. 
Its frequency indeed matches that of the bending mode of pure liquid water, as shown in Figure \ref{FTIR} (blue dotted curve). 

Considering the OH stretching region, as can be seen in Figure \ref{FTIR}, it appears to be rather smooth and broad, with no distinct features discernible. Nevertheless, when compared to liquid water (dotted blue curve) a significant amount of spectral weight is present at low frequencies, clearly indicating the occurrence of complex interactions between water molecules and a solid environment. In summary, this wide band incorporates all the OH stretching absorptions that originate from liquid water as well as hydrated Rb$_2$Ti$_2$O$_5$:(H$_2$O)$_x$ and RbHCO$_3$. 

\begin{figure}[h!]
\centering
\includegraphics[width=0.8\linewidth]{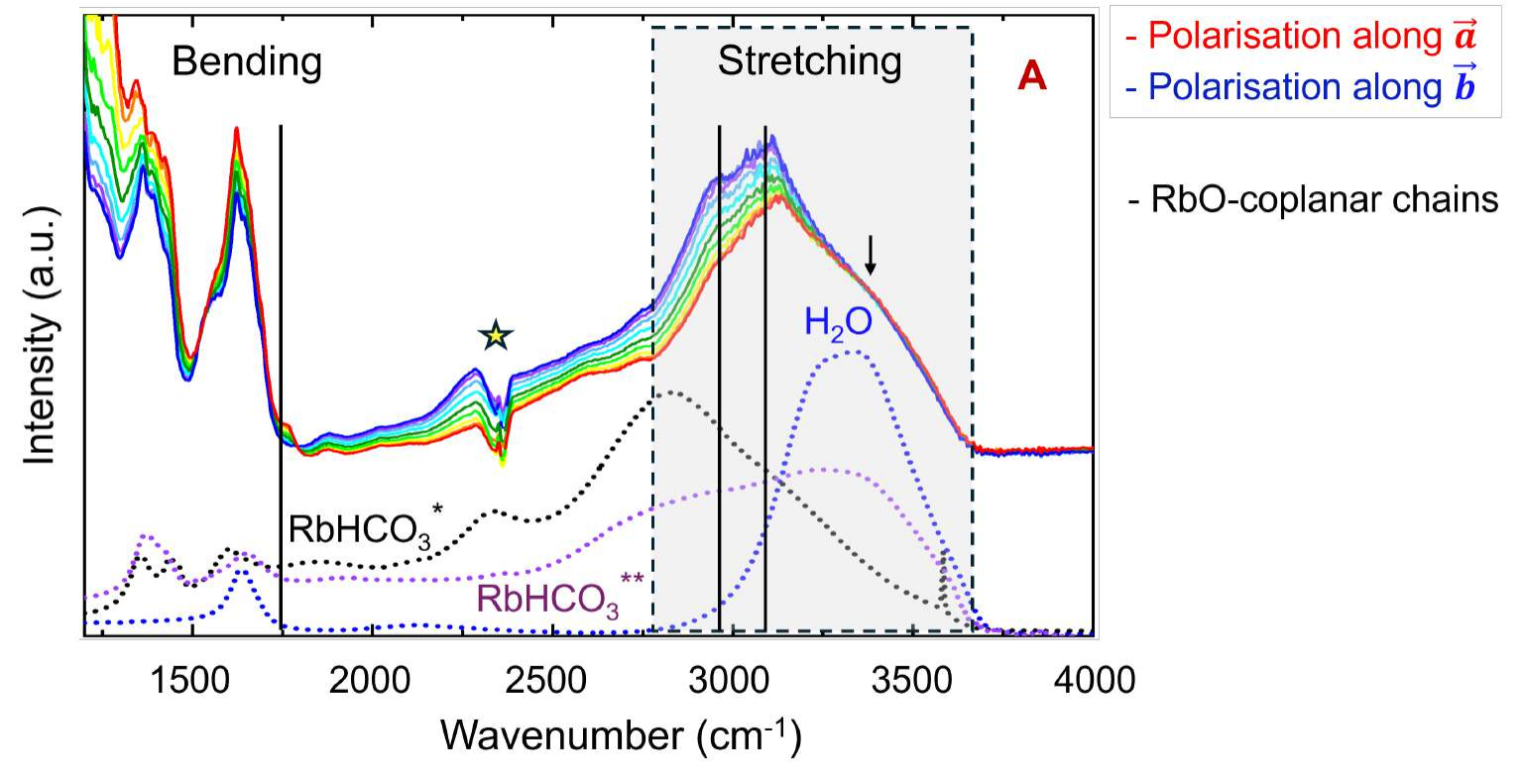}

\caption{\label{FTIR_H} A) Infrared spectra of a RTO single crystal hydrated during 16 h to ambient air. The RTO spectra were acquired as a function of polarization, ranging from $\vec{b}$ (blue solid curve) to $\vec{a}$ (red solid curve) in 10° increments. The star indicates the CO$_2$ antisymmetric stretching vibration band. The dotted curves correspond to the Attenuated Total Reflection (ATR) spectra of liquid water (blue), RbHCO${_3}^{*}$ (hydrated and carbonated rubidium dioxide powder, black) and RbHCO${_3}^{**}$ (hydrated solvated and carbonated rubidium dioxide, purple). The vertical solid lines indicate the DFT bending and stretching modes for the RbO-coplanar chains configuration (refer to Table \ref{tab:hyd_freq}). The vertical arrow indicates the spectral feature due to the stretching absorption of free water molecules.}
\end{figure}

\vspace{0.2cm}
Figure \ref{FTIR_H} instead shows the results as obtained for RTO crystals that have been hydrated for 16 hours in ambient air (at 34\% of humidity). 
Notably, when comparing the IR spectra of the highly hydrated samples to the less hydrated ones, other structures appear. 
In the stretching region (gray shadow frame) three distinct features can be discerned: two overlapped peaks and one shoulder (indicated by the black arrow). 
The shoulder is insensitive to polarization and confirms the presence of non-oriented water molecules in the system. 
On the contrary, the intensity of the two peaks at 2970 cm$^{-1}$ and 3120 cm$^{-1}$ depends on the polarization: in particular, the former has largest spectral weight along $\vec{b}$. 
Their frequencies remarkably match those we compute for the antisymmetric (AS) and symmetric (SS) stretching, respectively, in the RbO-coplanar water chains, as reported in Table \ref{tab:hyd_freq} (vertical solid black lines in Figure \ref{FTIR_H}). 
In particular, the peak at 2970 cm$^{-1}$ matches the anti-symmetric modes of the water chains that vibrate at 2982 cm$^{-1}$, that DFT predicts to be polarized along $\vec{b}$, in perfect agreement with the experiment.  
Moreover, the fact that the anti-symmetric stretching mode is lower in frequency than the symmetric one as also foreseen by the theory, reveals the strong interaction between the ordered water chains and the structural motif of the Rb$_2$Ti$_2$O$_5$ (001) surface. 
Similarly, in the bending region, the spectrum shows a peak at around 1750 cm$^{-1}$, polarized along $\vec{a}$.
Our calculations predict the bending mode of water chains at the (001) surface to vibrate at 1729 cm$^{-1}$, totally polarized along $\vec{a}$.
Therefore, both frequencies and the dependence upon the polarized light indicate without ambiguity the presence of self-organized water molecules that align on the Rb$_2$Ti$_2$O$_5$ (001) surfaces along $\vec{b}$ (see Figure \ref{fig:hyd.img}, right).

At last, it is worth noting that a new band at 2350 cm$^{-1}$ (see Figure \ref{FTIR_H}) exhibits a strong polarization along $\vec{b}$ (whose estimation is affected by the non-compensated absorption of the atmospheric CO$_2$ antisymmetric stretching). The energy of this band perfectly matches the energy of the asymmetric stretching in condensed CO$_2$, which suggests that CO$_2$ co-adsorbs with water on the Rb$_2$Ti$_2$O$_5$ (001) surface.

In summary, the measured IR spectra and their dependence upon light polarization are in excellent agreement with the DFT simulations of linear oriented chains of water molecules on the Rb$_2$Ti$_2$O$_5$ (001) surface, aligned along $\vec{b}$. The spectra also suggest the presence of non-oriented water in the sample, in conjunction with carbonate species.

\newpage
\subsection{\label{conduc} Ionic conductivity.}

In order to investigate the effect of this specific water organization on ionic conductivity, we carried out electrochemical impedance spectroscopy measurements (EIS) on three RTO single crystals that were cut respectively along $\vec{a}$, $\vec{b}$ and $\vec{c}$. The preparation of each sample is detailed in the Methods section. 

\begin{figure}[h!]
\begin{center}
    \includegraphics[scale=0.225]{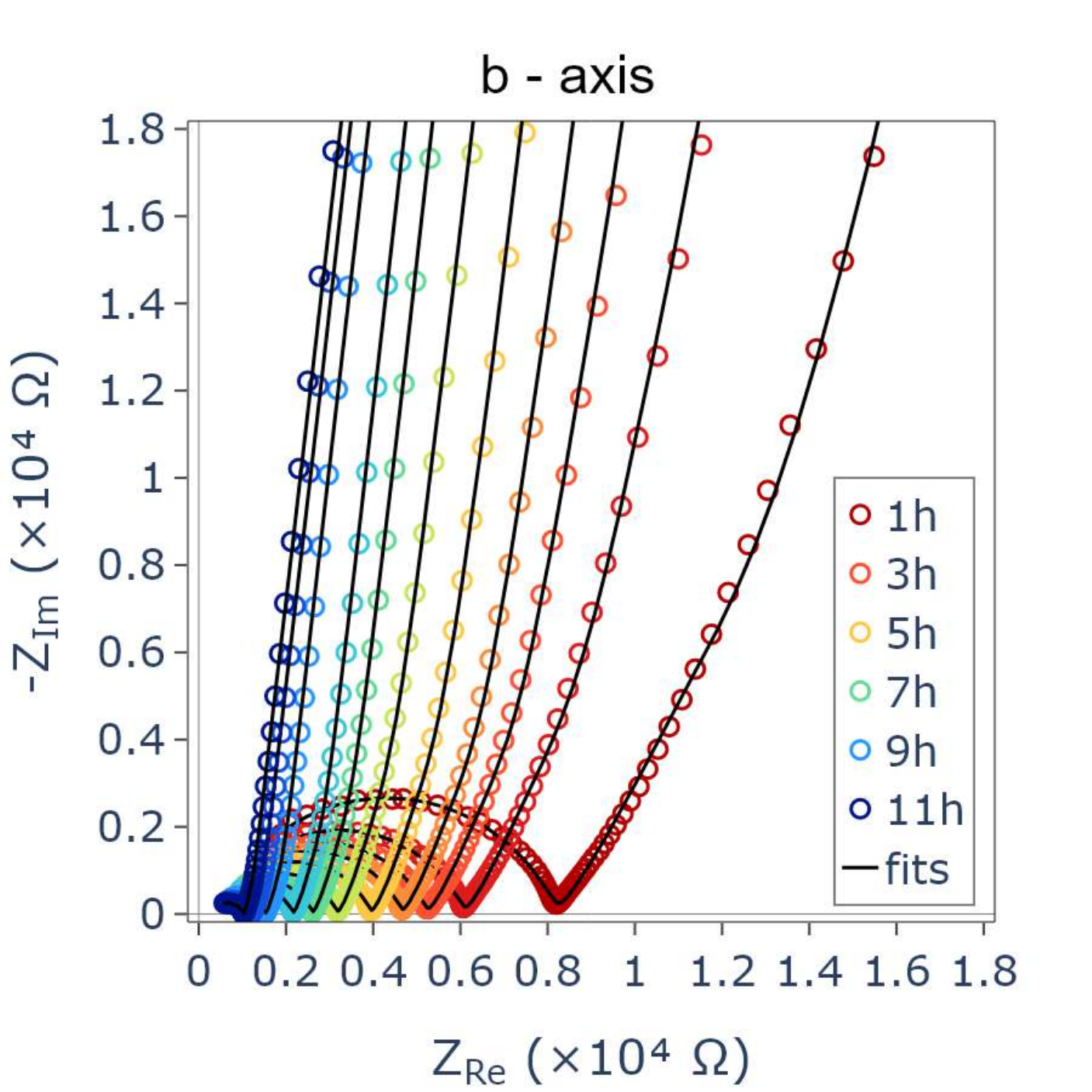}
\includegraphics[scale=0.2]{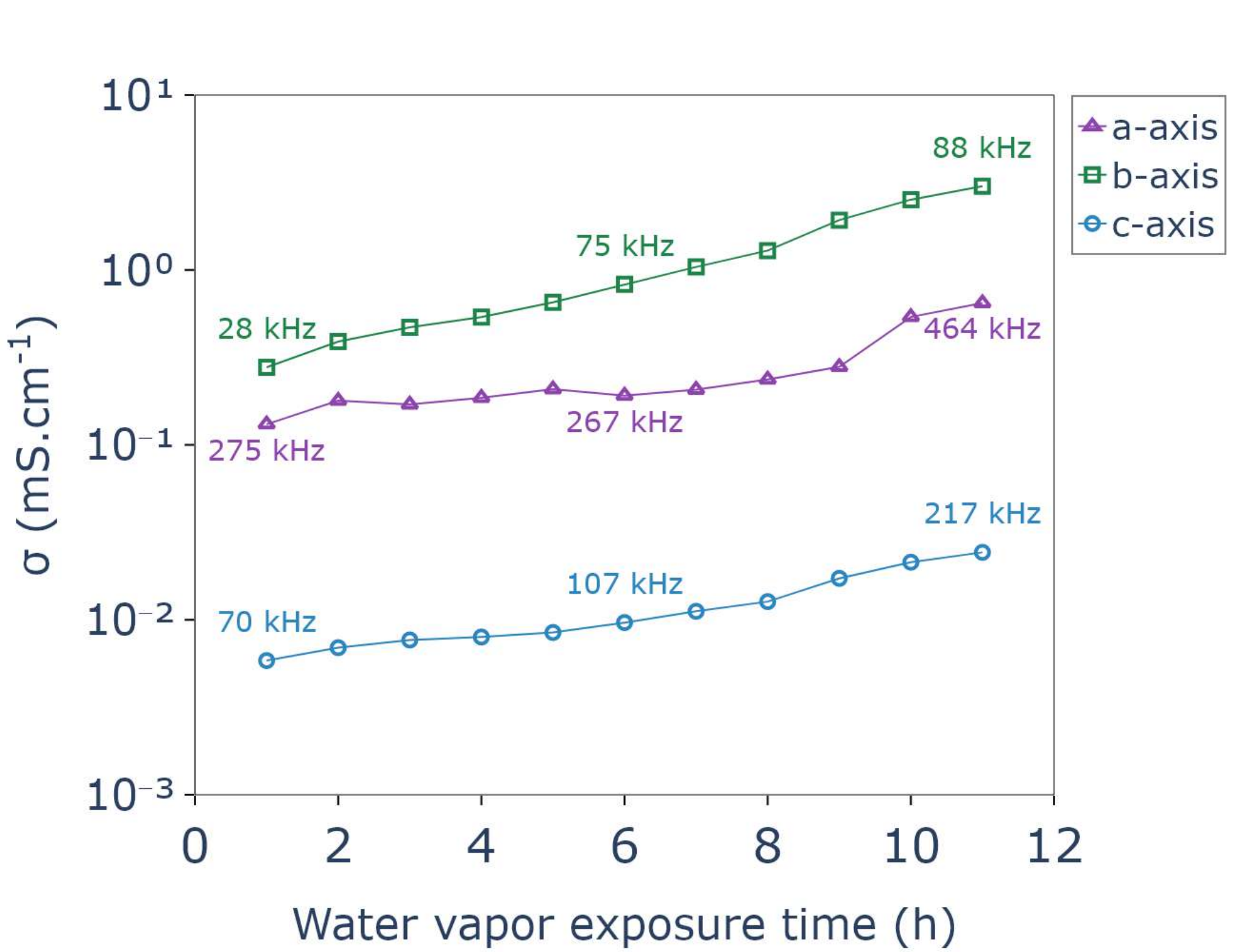}
\caption{\label{Nyquist1} Left : Nyquist plots of the impedance measured along $\vec{b}$ for a pristine RTO sample as function of time exposure to the laboratory atmosphere. The black lines indicate the fits obtained from the equivalent electrical circuit discussed in the text (Figure \ref{Nyquist_circuit}). Right: Conductivity of the sample for the three crystallographic directions, which were extracted from the Nyquist diagrams from the R$_{Bulk}$ value (see Supporting Information). The conductivity is highly anisotropic with about two orders of magnitude between $\vec{a}$ and $\vec{c}$ for 10 h of exposure time to the atmosphere. The characteristic frequency of the conductivity is labeled for exposure times of 1~h, 6~h and 11~h.}
\end{center}
\end{figure}

\begin{figure}[h!]
\begin{center}
    \includegraphics[scale=0.17]{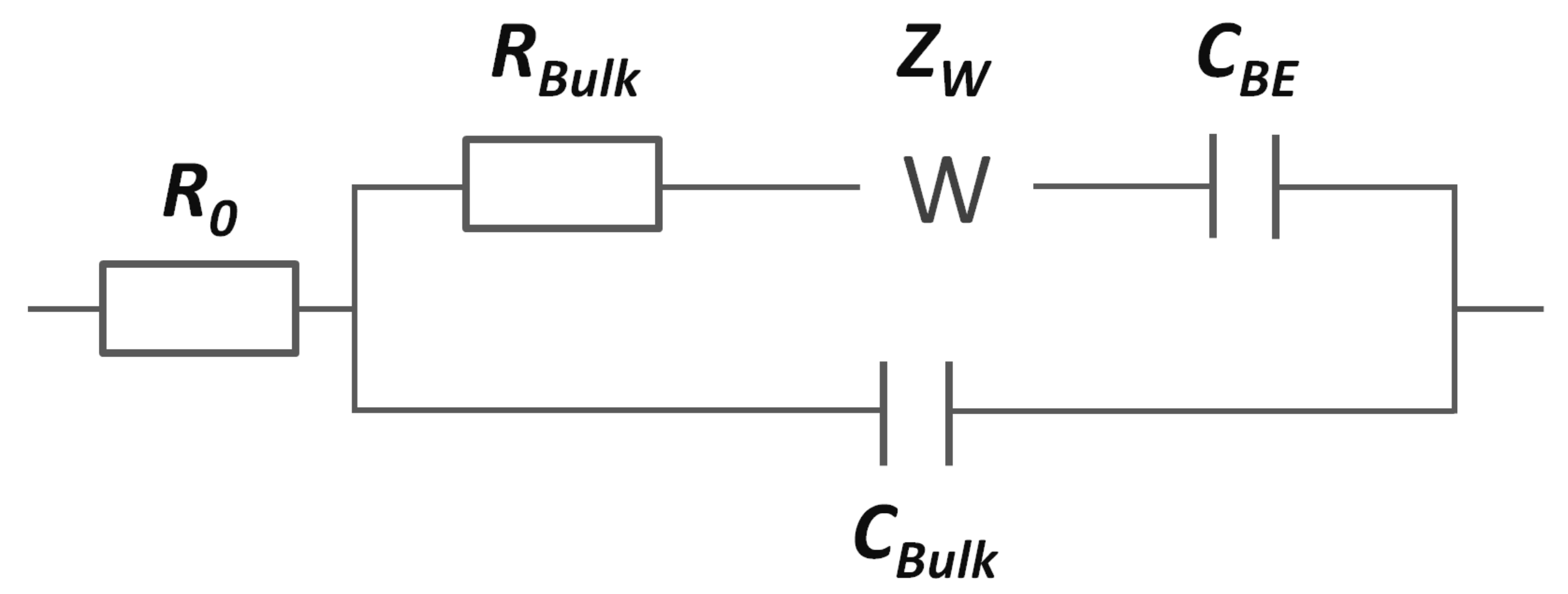}
\caption{\label{Nyquist_circuit} Equivalent electrical circuit for the RTO single crystal oriented along $\vec{b}$ used to model the EIS data. All capacitors used for the equivalent electrical circuit are Constant Phase Elements (CPE).}
\end{center}
\end{figure}

The Nyquist plots obtained from EIS measurements of the single crystal oriented along $\vec{b}$ at different exposure times are shown in the left panel of Figure \ref{Nyquist1}. The other Nyquist plots corresponding to the single crystals oriented along $\vec{a}$ and $\vec{c}$ are represented in the Supporting Information in Figure S9. All spectra display a semi-circle at high frequencies and a straight line at low frequencies. An equivalent electrical circuit was derived for each of the single crystal orientations to fit the collected EIS data \cite{huggins2002simple}. The equivalent electrical circuit for the single crystal oriented along $\vec{b}$ is detailed in Figure \ref{Nyquist_circuit}.

The high frequency region contains a resistance (R$_0$) that is associated to the electrical setup (the gold electrode deposition and wiring connections). The R$_0$ resistance is in series with the bulk resistance of the RTO single crystal (R$_{Bulk}$), which is parallel to the bulk capacitance (C$_{Bulk}$). The values of R$_{Bulk}$ and C$_{Bulk}$ obtained for the three orientations of the single crystals depending on the exposure time to the atmosphere are represented in Figure S10 in the Supporting Information.

The mid-frequency region is the most affected by the crystallographic orientation of the single crystal. This region is related to the diffusion and is strongly impacted by the tortuosity within the material, which is lower for the single crystal oriented along $\vec{b}$ compared to those oriented along $\vec{a}$ and $\vec{c}$. Regarding the diffusion along $\vec{b}$, the mid-frequency region can be associated to the conventional diffusion process represented by an open-circuit Warburg element (Z$_W$) in series with R$_{Bulk}$. In the presence of a stronger tortuosity effect, as observed along $\vec{a}$ and $\vec{c}$, the Warburg is replaced by a Transmission Line Model (TLM), located in series of R$_{Bulk}$. As a result, the diffusion pathways seem significantly impacted by the crystallographic orientation of the single crystal, and appears less complex along $\vec{b}$. The driving force for this diffusion to take place is the appearance of a water concentration gradient between the inner and outer part of the sample while exposed to the atmosphere. The decrease of this mid-frequency contribution in the EIS spectra as a function of exposure time is related to the kinetics of the water uptake and the achievement of equilibrium.

The low-frequency region corresponds to the interfacial response at the gold ion-blocking electrode, which is represented by a capacitor (C$_{BE}$) in series with R$_{Bulk}$, for all three crystallographic directions.

To study the effect of the crystallographic direction on the crystal conductivity, the bulk conductivity (left panel of Figure \ref{Nyquist1}) for all three orientations of the RTO single crystals depending on the exposure time to the laboratory atmosphere was retrieved by using the geometry of the crystal and the value of R$_{Bulk}$. A notable observation is the increase of the bulk conductivity for all three orientations of the single crystals with the increase of the exposure time to the atmosphere. Additionally, the characteristic frequency of the conductivity is in the range of 10$^4$~Hz and 10$^5$~Hz, suggesting the presence of proton conduction within the material.

Remarkably, the EIS results show a pronounced anisotropy in conductivity along the different crystallographic axes (left panel of Figure \ref{Nyquist1}). Specifically, the highest conductivity is along $\vec{b}$ with 0.28 mS/cm after 1~h of exposure time to the atmosphere, which is more than two orders of magnitude than that along $\vec{c}$, with a conductivity of 0.006 mS/cm. The anisotropy of the conductivity obtained from the RTO single crystal is consistent with DFT calculations, and points to Grotthuss mechanism for proton diffusion along the water chains. 
Furthermore, the conductivity continues to increase with prolonged exposure time to the atmosphere, reaching 3~mS/cm along $\vec{b}$ after 10~h. To the best of our knowledge, this value represents one of the highest protonic conductivities reported for solid crystalline oxides at room temperature.

\section{Discussion}
\label{Disc}

Both K$_2$Ti$_2$O$_5$ and Rb$_2$Ti$_2$O$_5$ are very hygroscopic materials, with high ionic conduction upon hydration.
Water incorporation gives rise to anisotropic ionic conductance with exceptionally high values of conductivity - up to 3 mS/cm at room temperature. 
The anisotropy in the conductance, which is preserved with increasing hydration, indicates that even in the highly hydrated samples - typically after 10 h of exposure, water incorporation is mostly anisotropic. 
The typical frequency (in the 10$^5$ Hz range) as extracted from the  analysis of electro-chemical impedance measurements is in favor of proton conduction and the evolution of this frequency with exposure to water suggests that the increase in the conductivity is due to a decrease in the proton scattering time.

At the nm scale, the hydration of both K$_2$Ti$_2$O$_5$ and Rb$_2$Ti$_2$O$_5$ results in the formation of heterogeneous regions, as revealed by scanning electron microscopy. 
The images consist of elongated contrasts along $\vec{b}$ that are split from each other by an almost constant distance ($\simeq 40-50$~nm) in the (010) plane.
Therefore these heterogeneous regions show a long-range order that is reminiscent of the crystal axes of the dry crystals. 
Although a reasonable hypothesis could be the presence of regions with variable hydration degrees, their precise compositions remains still to be determined.

First-principles simulations identify the most stable configuration for an isolated water molecule in the bulk crystals.
Such a configuration is prone to
form ordered water chains, aligned along $\vec{b}$ and strongly bound to the host crystal.
Their presence in the bulk would 
lead to a noticeable expansion of the crystal, mostly along $\vec{c}$, which is not detected in the x-ray diffraction experiments. 
However, the hydrated crystals are also predicted to be unstable against the formation of hydrated (001) surfaces, \cite{Benas2025} with water organized in chains (as in the bulk) on the surface planes. 
Therefore, a plausible explanation consists in the formation of heterogeneous regions, with interfaces between water (and other chemical species) and the alkali titanates. The water chains at the surface would be not altered, as they are very stable.   
These hydrated surfaces could also lead to anisotropic ion conductivity with preferred conductivity along $\vec{b}$. 
It is rather likely that the hydrated samples as used for the conductivity measurements are indeed made of hydrated proton-conducting surfaces consisting of self-organized one-dimensional water chains.

These predictions are well supported by IR absorption spectroscopy. 
Indeed, for the lightly hydrated compound, both frequencies and dependence on the polarization orientation of the bending mode reveal that water is adsorbed in ordered configurations (with a prevalence of RbO-coplanar configurations), although some randomly oriented water (presumably at the surface) is also detected. 
For the sample hydrated 16 hours under air exposure, the vibrational modes from RbO-coplanar are still present -- and even reinforced -- with an additional contribution, presumably from hydroxides (RbOH) and carbonates ($\mbox{Rb}_2\mbox{CO}_3$).

More systematic electron microscopy studies remain to be done in order to study the evolution of these heterogeneous contrasts as a function of the temperature and water content. Atomic force microscopy and/or environmental transmission electron microscopy could be the methods of choice for performing these analyses. In addition, molecular dynamics to come will help us understand the migration of protons or proton vacancies along the one-dimensional water chains at the (001) surfaces.

 The atomic-scale mechanisms for creation of the carriers and their mobility will therefore be the object of future studies. 
 In particular, it is necessary to understand the Grotthuss mechanism that takes place in these materials and the nature of the involved defects.
 To the best of our knowledge, this mechanism of incorporation of water molecules into self-organized one-dimensional chains is unique, and distinct from water incorporation in other lamellar Ti-oxides that exhibit proton conduction. For instance, in Ref.\cite{Kang2020}, proton conduction is evidenced in a hydrated titanate layered structure both in the inter-layer space and with the terminal oxide of the titanate framework, but without specific organization of the water molecules.

\section{Conclusion}

Simulations within the Density Functional Theory in conjunction with infra-red spectroscopy suggest an ordered arrangement of water molecules in Rb$_2$Ti$_2$O$_5$, with RbO-coplanar water molecules hydrogen bound to the apical O of the host crystal. 
At very low water concentration, the molecules form self-organized one-dimensional (1D) face-to-face double chains that are thermodynamically very stable.

At higher hydration, water chains are shown to organize at (001) surfaces created within the bulk material. These chains might thus constitute a suitable medium to promote anisotropic proton conduction or proton-vacancy conduction,  which is attested by complex impedance measurements that display high super-ionic conductivity (up to 3 mS$\cdot$cm$^{-1}$) along $\vec{b}$. 
Electron microscopy observations show that heterogeneous volume contrasts arranged in an orderly manner along $\vec{b}$ are observed in the case of hydrated K$_2$Ti$_2$O$_5$ and Rb$_2$Ti$_2$O$_5$ single crystals.
Notably, the conductivity anisotropy faithfully follows the structural anisotropy of Rb$_2$Ti$_2$O$_5$, thus indicating a correlation between the organization of the heterogeneous contrasts and the ion conductivity. 
The high scattering frequency is a signature for proton conduction, and the increase in frequency observed with increased hydration denotes that the very mobility of protons increases with water content.

All these observations and simulations therefore point to the self-organization of guest water molecules into one-dimensional patterns, in the bulk at the onset of water adsorption, then at hydrated (001) surfaces that spontaneously form in highly hydrated samples.
This conceptual frame provides a consistent explanation for the remarkably high anisotropic ionic conductivity in RTO and KTO crystals, unraveling a peculiar water incorporation that is distinct from other hygroscopic oxides.  
Hydrated KTO and RTO are therefore characterized by water self-organization into proton-conducting chains, which enables them for applications in energy-storage devices. More work remains to be done in order to understand the specific organization within the 1D-water organization, the way protons release and their dynamics, which presumably relates to a Grotthuss-like mechanism.

\section{Methods Section}
\textit{Sample preparation and hydration}\\

Crystals of $\mbox{M}_2\mbox{Ti}_2\mbox{O}_5$ were prepared using a protocol described in the supplemental material of \cite{Federicci_PRM_2017}. Powders of $\mbox{TiO}_2$ and RbNO$_3$ or KNO$_3$ precursors were mixed, ground and pressed into a pellet and then heated up in a furnace, as described in \cite{Federicci_PRM_2017}. After a slow cooling down, the crystalline structure was characterized by X-ray diffraction and found consistent with the one established in \cite{Federicci_AC_2017}, whose space group is \textit{C/2m} and the lattice parameters are the following: a$=$11.34 \AA, b$=$3.82 \AA, c$=$7.00 \AA, $\beta=$100.24° and $\alpha=\gamma=$90°.

Given the hygroscopic nature of the samples evidenced by previous studies \cite{desousacoutinho_SSI_2021, Meziani_2024}, a particular care was given to control the water vapor (and CO$_2$) exposure. The samples were stored inside a glove box under N$_2$ atmosphere immediately after the synthesis and whenever they were not in use and then put into hermetically sealed boxes (HSB), themselves placed inside a second HSB for transportation. They were exposed to air only during the minimal necessary time, as described in the experimental sections. In particular the furnace was opened at about 200\,°C during the cooling down after the synthesis and the crucibles were immediately placed into the glove box, entailing a minimal exposition time to the laboratory atmosphere (about 2 min). The crystals were consequently extracted from the crucible inside the glove box. Therefore in the present study we consider as "pristine" a sample that has been kept either inside the glove box or inside an hermetically sealed plastic box under N$_2$ or Ar atmosphere. IR absorption measurements  and TG analysis presented in \cite{Meziani_2024} indeed show that this procedure leads to minimal absorption of water and minimal formation of carbonates in RTO and KTO powders but with a very different hydration velocity. RTO powders were indeed found to hydrate 10 times faster than KTO powders \cite{Meziani_2024}.\\

\textit{Electron microscopy experiments}\\

\begin{figure*}[h!]
\includegraphics[scale=0.18]{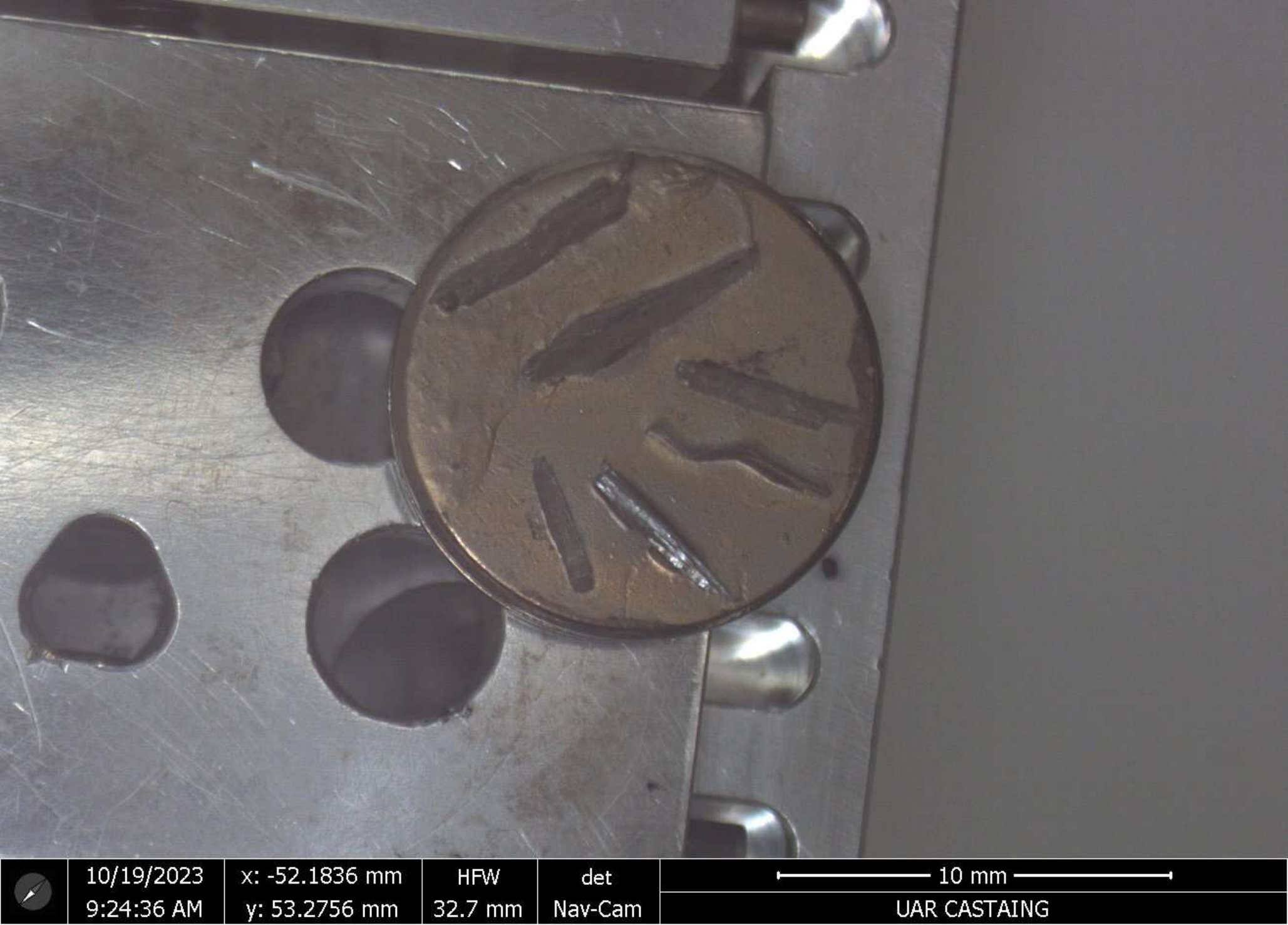}%
\hfill
\includegraphics[scale=0.460]{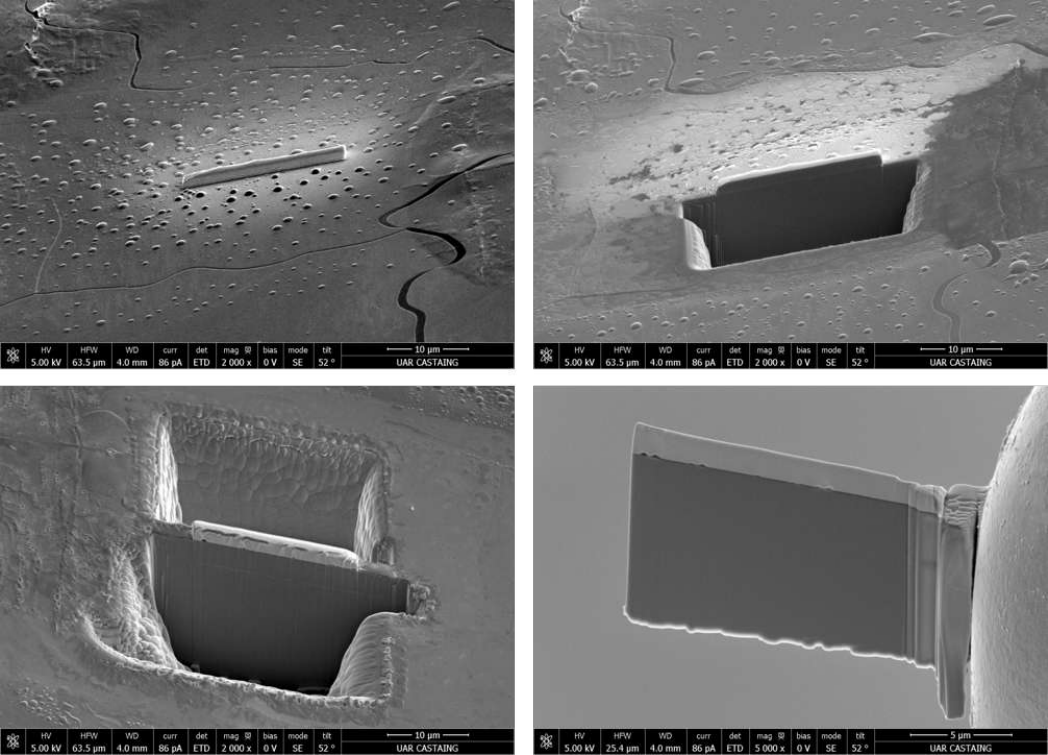}
\caption{\label{fig9MEB} Left: Image of the stub inside the FIB chamber. Six single crystals are glued to the stub with silver paste that was applied over the whole surface of the stub. Right: Illustration of the different steps of the FIB milling process. a) corresponds to the sample after carbon and platinum deposition on a selected area in order to protect the sample surface during milling. b) and c) picture the sample respectively after the Ga-ion milling on one side and on the other side of the carbon and platinum deposit.  d) is a picture of the extracted platelet with suitable orientation.  }
\end{figure*}

As described above, crystals of RTO and KTO were removed from the synthesis furnace at 200°C and transferred into a glove box under Ar atmosphere within less than two minutes. Single crystals were then extracted and glued with silver paint onto microscope sample holders (stubs) inside the glove box (see Figure \ref{fig9MEB}). The stubs were then placed into hermetically sealed boxes (HSB), in order to be transported to the site of the electron microscope setup.

Immediately after opening the HSB, carbon was deposited (by sputtering and carbon evaporation Leica EM ACE600 I) on the single crystals of RTO and KTO to prepare them for the Ga-focused ion beam (FIB). This operation was required given the very low electronic conduction of the materials. Then the single crystals were transported from the carbon deposition machine directly to a focused-ion-beam-equipped scanning electron microscope (FIB-SEM) and the total time for this transfer operation was about 3 min. After placing the stubs in the FIB chamber, a protective layer of carbon and platinum was desposited on top of the area of interest, before starting  Ga-milling of the samples.  The different steps for milling are illustrated in Figure \ref{fig9MEB} (right). \\

\textit{IR spectroscopy measurements}\\

Mid-infrared spectroscopy measurements were conducted in ambient air at room temperature using a Bruker ALPHA spectrometer operating in transmission geometry and equipped with a deuterated triglycine sulfate (DTGS) detector, a KBr beam splitter and a global source. The recorded energy range was 700$-$4000 cm$^{-1}$. RTO monocrystals (of 0.54 mm thick, 3 mm long, and 0.6 mm wide) were characterized using polarization-dependent measurements. RTO crystals were aligned along $\vec{b}$, and the polarizer’s orientation was varied for each measurement. Polarization at 0° corresponds to alignment along $\vec{b}$, while polarization at 90° corresponds to alignment along $\vec{a}$. For accuracy, a reference spectrum was recorded by measuring the sample holder (with a 2 mm aperture) under identical ambient conditions. All infrared spectra have been corrected using the reference spectrum mentioned above.
RbOH powder sample was exposed to air, which promotes carbonation, leading to the expected formation of RbHCO$_3$  with different hydration content. The carbonated RbOH and distilled water, were measured using a diamond ATR system set-up on a Bruker ALPHA.\\

\newpage
\textit{Ionic conductivity measurements}\\

Three pristine single crystals issued from the same synthesis batch were isolated and placed into a micro-tube along different directions: $\vec{a}$, $\vec{b}$, and $\vec{c}$. The micro-tube was then filled with transparent epoxy resin (Resinpro, iCrystal). Once the resin was solidified at 100°C, the assembly was cut using a diamond-wheel saw, resulting slices of the sample embedded in resin with a thickness of 5.5~mm, 5.2~mm and 5.9~mm for direction along $\vec{a}$, $\vec{b}$ and $\vec{c}$, respectively. Subsequently a thin film of 30~nm of gold (Au) was deposited onto both sides of the three samples by evaporation deposition, and silver wires were attached on each side of the sample using silver conductive lacquer (RS Components) to establish electrical contacts with the potentiostat alligator clips. The first side of the sample served as a working electrode (WE) with a surface of 0.67~mm$^2$, 5.5~mm$^2$ and 2.7~mm$^2$ for $\vec{a}$, $\vec{b}$ and $\vec{c}$, respectively. While the other side served as a counter electrode (CE), resulting in a 2-electrode system. EIS measurements were performed using a potentiostat (Biologic SP 300). In solid-state electrolytes, EIS allows the evaluation of resistances present within the solid material, which can originate from both the grains and grain boundaries of the solid electrolytes \cite{LARFAILLOU2016139,Vadhva_2021,Lazanas_2023}. Thereby, EIS can be effectively used at high frequencies to extract the intrinsic conductivity of the electrolyte. In this work, EIS measurements were conducted at open circuit potential (OCP) within a frequency range from 3 MHz to 1 Hz along $\vec{a}$, from 3 MHz to 0.1 Hz along $\vec{c}$ and from 7 MHz to 0.3 Hz along $\vec{b}$. The amplitude for all measurements was fixed at 10 mV, and 12 points per decade to extract and compare the ionic conductivity of the samples along their various directions and at different durations of exposure to around 40-50\% relative humidity. The fitting of the Nyquist diagrams was performed using the Yappari software \cite{dragoe2023yappari}. \\

\textit{Ab-initio simulations }\\

Density Functional Theory and Density Function Perturbed Theory calculations were both performed using the Quantum ESPRESSO package \cite{Giannozzi2017}. The exchange-correlation functional was modeled using the revised version for solids (PBEsol) of the Perdew-Burke-Ernzerhof functional \cite{Perdew1996,revPBE1998}. Optimized norm-conserving pseudo-potentials \cite{Hamann2013} were used to describe the interaction between ions and valence electrons. A minimum energy cutoff of $1088$ eV ($\sim$ 80 Ry) was employed for the plane-wave basis set. The (001) surface was modeled using the slab construction described in \cite{Benas2025}. The water molecules were initially inserted into low-density regions of the crystal or by taking into account electrostatic considerations, and their positions were relaxed to minimize interatomic forces down to 10$^{-2}$ eV.\AA$^{-1}$. The stability of the hydration process $\mbox{Rb}_2\mbox{Ti}_2\mbox{O}_5+x\ \mbox{H}_2\mbox{O}^{(gas)}\longrightarrow\mbox{Rb}_2\mbox{Ti}_2\mbox{O}_5:\left(\mbox{H}_2\mbox{O}\right)_x$ was evaluated at T = 0 K for all the configurations by taking the total energy difference between the hydrated systems on one side, and the dry bulk system along with isolated water molecules on the other side. This energy difference defines the formation energy $\Delta E_f$, allowing the comparison between the bulk and surface systems. The perturbative calculations are carried out at the harmonic level following Ref. \cite{DFPTT}. Technical aspects, including the use of a reduced set of atomic displacements and the scaling of vibrational frequencies, are detailed in the Supplementary Material.

\medskip
\textbf{Supporting Information} \par 
Supporting Information is available from the Wiley Online Library or from the author.

\medskip
\textbf{Acknowledgments} \par 

This work was supported by the French Agence Nationale de la Recherche (ANR) for funding,
through the project MIMETIX (ANR 21-CE50-0035-02). The authors acknowledge financial support from the CNRS-CEA METSA French network (FR CNRS 3507) on the CLyM platform. The work at LPEM was partly supported by PSL University through a maturation project in the framework of the “Investissements d’avenir” ANR program.

\medskip

%
\bibliographystyle{MSP}
\bibliography{am2025_bib}

\begin{thebibliography}{10}
\providecommand{\url}[1]{\texttt{#1}}
\providecommand{\urlprefix}{URL }

\bibitem{Meng_CSR_2017}
X.~Meng, H.-N. Wang, S.-Y. Song, H.-J. Zhang,
\newblock \emph{Chem. Soc. Rev.} \textbf{2017}, \emph{46} 464.

\bibitem{FOP2021}
S.~Fop,
\newblock \emph{J. Mater. Chem. A} \textbf{2021}, \emph{9} 18836.

\bibitem{Xu_2021}
Y.~Xu, X.~Wu, X.~Ji,
\newblock \emph{Small Structures} \textbf{2021}, \emph{2}, 5 2000113.

\bibitem{HUANG_2023}
C.~Huang, W.~Zhang, W.~Zheng,
\newblock \emph{Energy Storage Materials} \textbf{2023}, \emph{61} 102913.

\bibitem{Grotthuss_1806}
C.~de~Grotthuss,
\newblock \emph{Ann. Chim.} \textbf{1806}, \emph{58} 54.

\bibitem{AGMON_1995}
N.~Agmon,
\newblock \emph{Chemical Physics Letters} \textbf{1995}, \emph{244}, 5 456.

\bibitem{Shimizu_2013}
G.~K.~H. Shimizu, J.~M. Taylor, S.~Kim,
\newblock \emph{Science} \textbf{2013}, \emph{341}, 6144 354.

\bibitem{Zhao_CM_2014}
X.~B. Xiang~Zhao, Chengyu~Mao, P.~Feng,
\newblock \emph{Chem. Mat.} \textbf{2014}, \emph{26}.

\bibitem{Chandra_JACS_2014}
S.~Chandra, T.~Kundu, S.~Kandambeth, R.~Babarao, Y.~Marathe, S.~M. Kunjir, R.~Banerjee,
\newblock \emph{J. Am. Chem. Soc.} \textbf{2014}, \emph{136} 6570.

\bibitem{YEUNG_CT_2024}
K.~L. Yeung, W.~Han,
\newblock \emph{Catalysis Today} \textbf{2014}, \emph{236} 182.

\bibitem{Sun_2018}
P.~Sun, R.~Ma, T.~Sasaki,
\newblock \emph{Chem. Sci.} \textbf{2018}, \emph{9} 33.

\bibitem{andersson_five_1960}
S.~Andersson, A.~D. Wadsley,
\newblock \emph{Nature} \textbf{1960}, \emph{187} 499.

\bibitem{Federicci_PRM_2017}
R.~Federicci, S.~Hol\'e, A.~F. Popa, L.~Brohan, B.~Baptiste, S.~Mercone, B.~Leridon,
\newblock \emph{Phys. Rev. Mater.} \textbf{2017}, \emph{1} 032001.

\bibitem{federicci_memory_2018}
R.~Federicci, S.~Holé, V.~Démery, B.~Leridon,
\newblock \emph{Journal of Applied Physics} \textbf{2018}, \emph{124}, 15 152104, \_eprint: https://pubs.aip.org/aip/jap/article-pdf/doi/10.1063/1.5036841/15216969/152104\_1\_online.pdf.

\bibitem{DESOUSACOUTINHO201972}
S.~{De Sousa Coutinho}, R.~Federicci, S.~Holé, B.~Leridon,
\newblock \emph{Solid State Ionics} \textbf{2019}, \emph{333} 72.

\bibitem{RANI2020126784}
R.~Rani, S.~{De Sousa Coutinho}, S.~Holé, B.~Leridon,
\newblock \emph{Materials Letters} \textbf{2020}, \emph{258} 126784.

\bibitem{desousacoutinho_SSI_2021}
S.~De~Sousa~Coutinho, D.~Bérardan, G.~Lang, R.~Federicci, P.~Giura, K.~Beneut, N.~Dragoe, S.~Holé, B.~Leridon,
\newblock \emph{Solid State Ionics} \textbf{2021}, \emph{364} 115630.

\bibitem{Meziani_2024}
N.~Meziani, M.~Parent, G.~Rousse, B.~Leridon, D.~Berardan, P.~Giura,
\newblock \emph{Inorganic Chemistry} \textbf{2024}, \emph{63}, 38 17513, pMID: 39225367.

\bibitem{Federicci_AC_2017}
R.~Federicci, B.~Baptiste, F.~Finocchi, F.~Popa, L.~Brohan, K.~Béneut, P.~Giura, G.~Rousse, A.~Descamps-Mandine, T.~Douillard, A.~Shukla, B.~Leridon,
\newblock \emph{Acta Crystallographica Section B} \textbf{2017}, \emph{73}, 6 1142.

\bibitem{xcrysden}
A.~Kokalj,
\newblock \emph{Journal of Molecular Graphics and Modelling} \textbf{1999}, \emph{Volume 17}, Issues 3–4.

\bibitem{Yoon2025}
C.~Yoon, H.~Han, Y.~Kim, H.-J. Shin,
\newblock \emph{Applied Surface Science} \textbf{2025}, \emph{689} 162507.

\bibitem{Benas2025}
G.~Benas, S.~De~Sousa~Coutinho, B.~Leridon, F.~Finocchi,
\newblock \emph{Phys. Chem. Chem. Phys.} \textbf{2025}, --.

\bibitem{Larvor}
C.~Larvor, B.~St{\"{o}}ger,
\newblock \emph{Acta Crystallographica Section E} \textbf{2017}, \emph{73}, 7 975.

\bibitem{odin}
C.~Odin,
\newblock \emph{Magnetic Resonance in Chemistry} \textbf{2004}, \emph{42}, 4 381.

\bibitem{munst2021infrared}
M.~G. M{\"u}nst, M.~On{\v{c}}{\'a}k, M.~K. Beyer, C.~van~der Linde,
\newblock \emph{The Journal of Chemical Physics} \textbf{2021}, \emph{154}, 8.

\bibitem{huggins2002simple}
R.~A. Huggins,
\newblock \emph{Ionics} \textbf{2002}, \emph{8}, 3 300.

\bibitem{Kang2020}
S.~Kang, A.~Singh, K.~G. Reeves, J.-C. Badot, S.~Durand-Vidal, C.~Legein, M.~Body, O.~Dubrunfaut, O.~J. Borkiewicz, B.~Tremblay, C.~Laberty-Robert, D.~Dambournet,
\newblock \emph{Chemistry of Materials} \textbf{2020}, \emph{32}, 21 9458.

\bibitem{LARFAILLOU2016139}
S.~Larfaillou, D.~Guy-Bouyssou, F.~{le Cras}, S.~Franger,
\newblock \emph{Journal of Power Sources} \textbf{2016}, \emph{319} 139.

\bibitem{Vadhva_2021}
P.~Vadhva, J.~Hu, M.~J. Johnson, R.~Stocker, M.~Braglia, D.~J.~L. Brett, A.~J.~E. Rettie,
\newblock \emph{ChemElectroChem} \textbf{2021}, \emph{8}, 11 1930.

\bibitem{Lazanas_2023}
A.~C. Lazanas, M.~I. Prodromidis,
\newblock \emph{ACS Measurement Science Au} \textbf{2023}, \emph{3}.

\bibitem{dragoe2023yappari}
N.~Dragoe,
\newblock \emph{Materials Lab} \textbf{2023}, \emph{2}.

\bibitem{Giannozzi2017}
P.~Giannozzi, O.~Andreussi, T.~Brumme, O.~Bunau, M.~B. Nardelli, M.~Calandra, R.~Car, C.~Cavazzoni, D.~Ceresoli, M.~Cococcioni, N.~Colonna, I.~Carnimeo, A.~D. Corso, S.~de~Gironcoli, P.~Delugas, R.~A. DiStasio, A.~Ferretti, A.~Floris, G.~Fratesi, G.~Fugallo, R.~Gebauer, U.~Gerstmann, F.~Giustino, T.~Gorni, J.~Jia, M.~Kawamura, H.-Y. Ko, A.~Kokalj, E.~Küçükbenli, M.~Lazzeri, M.~Marsili, N.~Marzari, F.~Mauri, N.~L. Nguyen, H.-V. Nguyen, A.~O. de-la Roza, L.~Paulatto, S.~Poncé, D.~Rocca, R.~Sabatini, B.~Santra, M.~Schlipf, A.~P. Seitsonen, A.~Smogunov, I.~Timrov, T.~Thonhauser, P.~Umari, N.~Vast, X.~Wu, S.~Baroni,
\newblock \emph{Journal of Physics: Condensed Matter} \textbf{2017}, \emph{29}, 46 465901.

\bibitem{Perdew1996}
J.~P. Perdew, K.~Burke, M.~Ernzerhof,
\newblock \emph{Phys. Rev. Lett.} \textbf{1996}, \emph{77} 3865.

\bibitem{revPBE1998}
Y.~Zhang, W.~Yang,
\newblock \emph{Phys. Rev. Lett.} \textbf{1998}, \emph{80} 890.

\bibitem{Hamann2013}
D.~R. Hamann,
\newblock \emph{Phys. Rev. B} \textbf{2013}, \emph{88} 085117.

\bibitem{DFPTT}
S.~Baroni, S.~de~Gironcoli, A.~Dal~Corso, P.~Giannozzi,
\newblock \emph{Rev. Mod. Phys.} \textbf{2001}, \emph{73} 515.

\bibitem{Shima}
T.~Shimanouchi,
\newblock Tables of molecular vibrational frequencies,
\newblock National Bureau of Standards, \textbf{1972}.

\bibitem{Kokalj2003}
A.~Kokalj,
\newblock \emph{Comp. Mater. Sci.} \textbf{2003}, \emph{28} 155.

\end{thebibliography}








\newpage
\clearpage
\appendix
\renewcommand{\thefigure}{S\arabic{figure}}
\renewcommand{\thetable}{S\arabic{table}}
\renewcommand{\thesection}{\arabic{section}}
\renewcommand{\theequation}{S\arabic{equation}}
\setcounter{figure}{0}
\setcounter{table}{0}
\setcounter{equation}{0}
\setcounter{section}{0}

\section*{Supplementary Information : One-dimensional self-organization of water molecules in proton conducting Andersson-Wadsley titanates.}

\pagestyle{fancy}
\fancyhf{}

\fancyhead[L]{\textbf{Supplementary Information}}
\fancyhead[R]{\thepage}

\renewcommand{\headrulewidth}{0.4pt}


\author{Mathilde Arnaud $^\#$, Guillaume Benas $^\#$, Narimane Meziani $^\#$, Armel Descamps-Mandine,  Claudie Josse, Sophia Akkari, Thierry Douillard, Sofia de Sousa Coutinho,  Mélanie De Vos, Stéphane Holé, Gwenaëlle Rousse, Rémi Federicci, Sylvain Franger, Paola Giura, Fabio Finocchi and Brigitte Leridon *}

\section{Electron microscopy observations}

\begin{figure}[h!]
\begin{center}
\includegraphics[scale=0.4]{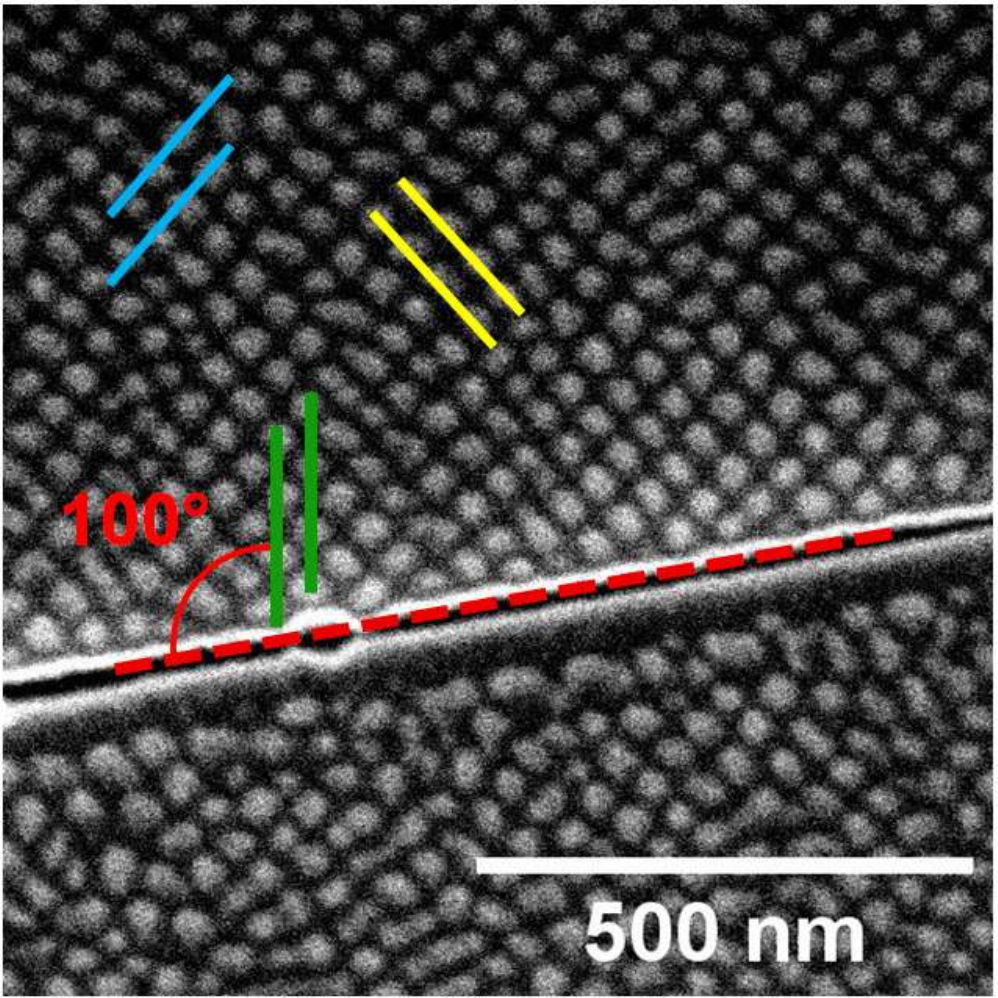}
    \includegraphics[scale=0.4]{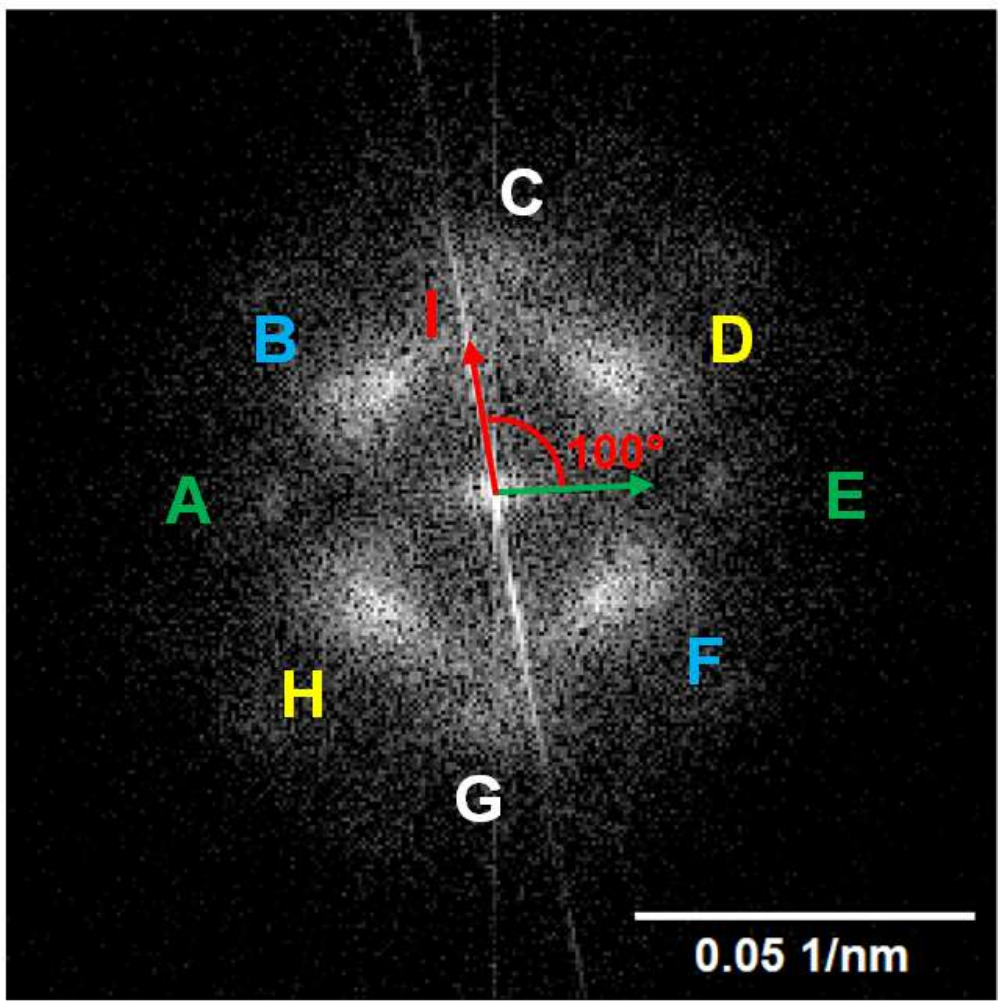}
\caption{\label{fig10MEB} Left panel: typical image observed on lightly hydrated RTO1b sample. Right panel: Fast Fourier Transform of the ordered white dot pattern observed in left panel.  This FT shows a superimposition of a quadratic network where: AE=CG, BF=DH, AE/BF $\approx\sqrt{2}$ and angles $\widehat{AOB}$=$\widehat{BOC}$=45° (with O as the center of the Fourier Transform). The green dash line represents $\vec{c}$ crystallographic orientation, orientated at 100° from the cleaving plane, which is represented by the red straight line. This shows that $\vec{c}$ in RTO matches with the orientation of the ordered white dot pattern (AE). The measurements were done with the \textit{ImageJ} software.}
\end{center}
\end{figure}

Figure \ref{fig10MEB} shows a typical SEM image taken on RTO1b as well as its Fourier transform. A long-range quasi-quadratic network is clearly visible. 

\begin{figure}[h!]
\begin{center}
  \includegraphics[width=0.4\linewidth]{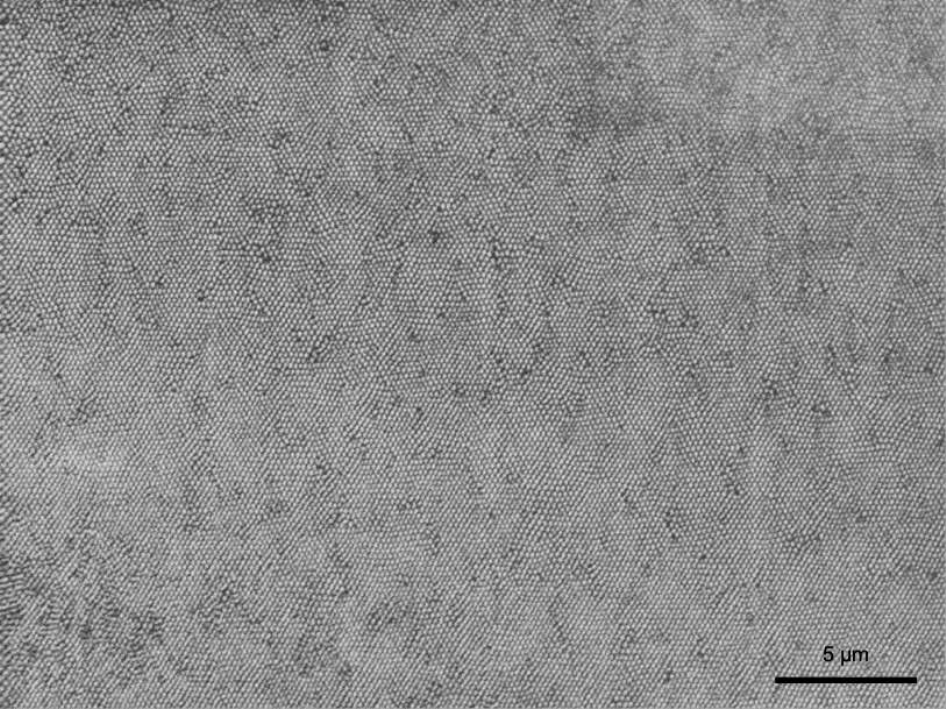}%
\includegraphics[width=0.3\linewidth]{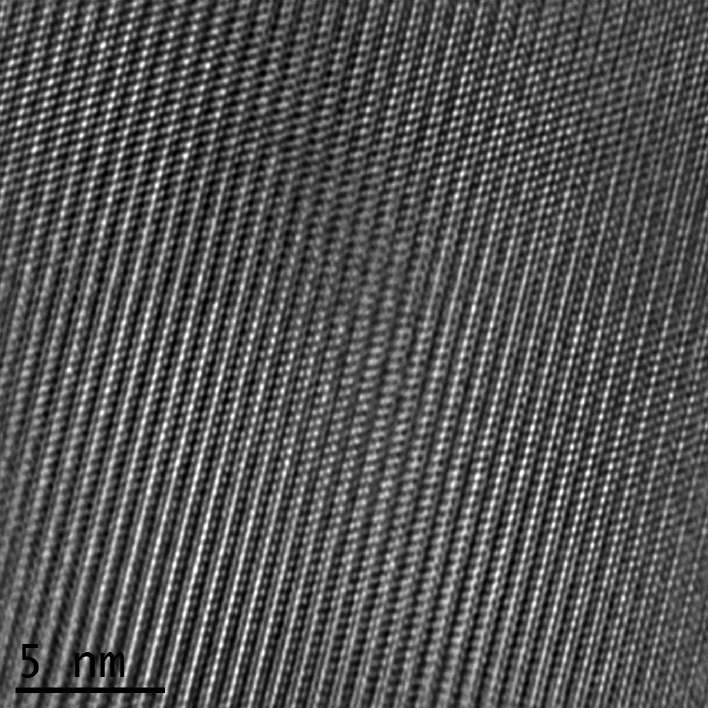}%
\includegraphics[width=0.3\linewidth]{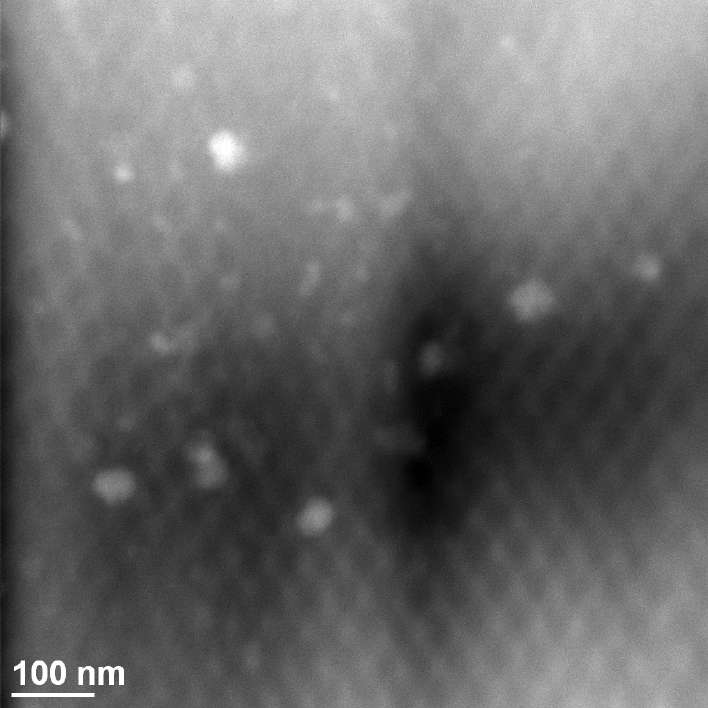}%
\caption{\label{LYON} Left: Low energy SEM observation of the ordered white dot pattern in a RTO platelet oriented perpendicular to $\vec{b}$ (RTO2b), with a local hexagonal symmetry (see Figure \ref{RTO2b_FFT_MEB}). Center: HRTEM observation of the same RTO platelet, confirming the existence of a well-preserved atomic network, corresponding to the RTO structure. Right: Observation of the ordered white dot pattern in the same RTO platelet, with a local hexagonal symmetry, using a STEM-HAADF microscope. This technique is mainly sensitive to chemical contrast, thus pointing to a modulation of the hosting network. }  
\end{center}
\end{figure}

\begin{figure}[h!]
\begin{center}
\includegraphics[scale=0.4]{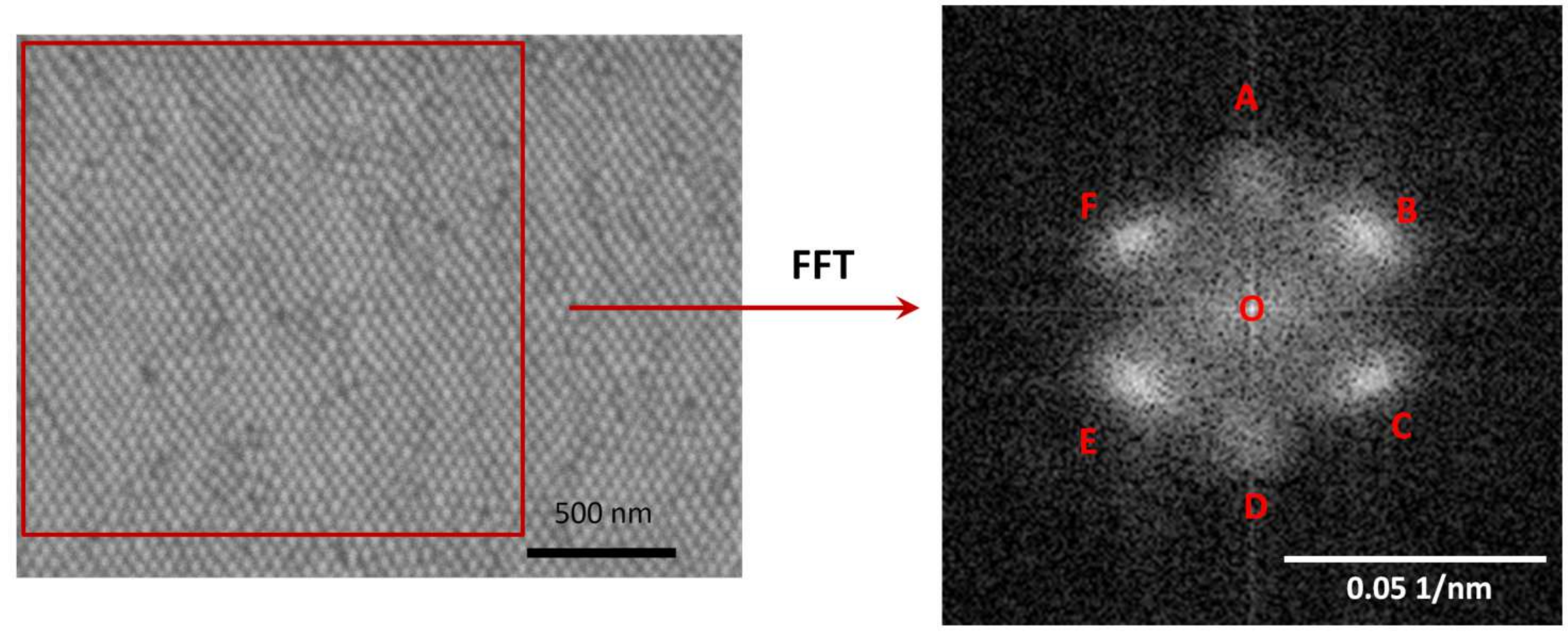}
\caption{\label{RTO2b_FFT_MEB} Left panel: A zoom from the left panel in Figure \ref{LYON} of the lightly hydrated RTO2b sample. Right panel: Fast Fourier Transform of the ordered white dot pattern observed in left panel.  This FT shows a hexagonal symmetry network where: AD$\approx$BE$\approx$CF and angles $\widehat{AOB}$=$\widehat{BOC}$=$\widehat{COD}$=60° (with O as the center of the Fourier Transform). The measurements were done with the \textit{ImageJ} software.}

\end{center}
\end{figure}

\clearpage

\section{DFT Calculations}

\subsection{IR spectra calculations}
\underline{Selected atomic vibrations:}
To reduce the computational cost of the DFPT calculations  performed at the harmonic level, only the displacements of a limited number of atoms are considered. Table~\ref{tab:selected_atoms} reports the vibrational frequencies of a single water molecule adsorbed in the bulk for the RbO-coplanar and Bridging configurations, obtained by considering either only the atomic displacements of the water molecule or additionally including those of its first-neighbor atoms. Here the frequencies are not scaled (see below). The resulting frequencies agree within 1~cm$^{-1}$, and the corresponding atomic displacement patterns are identical. On this basis, all perturbative calculations in this study are performed at the harmonic level by restricting the atomic displacements to the water molecules only. To reduce the offset arising from the DFT during frequency evaluation, a scaling procedure is presented below.

\begin{table}[h!]
    \centering
    \begin{tabular}{|c||c|c||c|c|}
    \hline
        Modes & \multicolumn{2}{|c||}{RbO-coplanar} & \multicolumn{2}{|c|}{Bridging} \\ \hline
        Selected atoms  &  H$_2$O  & H$_2$O + 1st neighbors &  H$_2$O  & H$_2$O + 1st neighbors \\ \hline \hline
        Bending &   1658.05   &   1657.99 & 1505.32  & 1505.56   \\ \hline
        Stretching & 2827.44 & 2827.50 & 2834.53 & 2834.34 \\ \hline
        Stretching  & 2932.22 & 2932.22  & 2978.66 & 2978.69 \\ \hline 
    \end{tabular}
    \caption{Vibrational frequencies (cm$^{-1}$) of a water molecule adsorbed in the bulk in the  "RbO-coplanar" and "Bridging" configurations. Calculations were performed considering either displacements of H$_2$O only or displacements of H$_2$O and its first neighboring atoms. Here the frequencies are not scaled (see below).}
    \label{tab:selected_atoms}
\end{table}

\noindent\underline{Scaling of the vibrational frequencies:} to minimize the systematic offset inherent to DFT calculations, the vibrational frequencies of the isolated molecule are aligned with the experimental values \cite{Shima}. The deviations between the theoretical frequencies obtained from DFPT at the harmonic level and the experimental values are quantified as ratios for the bending and stretching modes of the water molecule and are reported in the Table \ref{tab:renorm}. Since the bending mode already matches the experiment, it is not included in the scaling procedure. Accordingly, all frequencies reported in this study are scaled by a factor of 0.982.

\begin{table}[h!]
    \centering
    \begin{tabular}{|c||c|c||c|c|}
    \hline
        \rule{0pt}{2.5ex} \textbf{Modes} & \textbf{Exp.} & \textbf{Th.}   & $\mathbf{\delta =  \nu^{exp}/\nu^{th}}$ & $\mathbf{\bar{\delta}\cdot\nu}$ \\ \hline \hline
        Bending &  1595 & 1591 &   1.003 & / \\ \hline
        Sym. S. & 3657 & 3717 &  0.984  & 3650 \\ \hline
        Antisym. S. & 3756 & 3834   & 0.980  & 3765 \\ \hline 
    \end{tabular}
    \caption{Deviations between the experimental \cite{Shima} and theoretical vibrational frequencies of the isolated water molecule (in cm$^{-1}$) are reported, along with the frequencies scaled by a factor $\bar{\delta}=0.982$.}
    \label{tab:renorm}
\end{table}

\newpage
\subsection{Bulk adsorption sites for water molecule}
\begin{figure}[h!]
\centering
\includegraphics[width=0.8\linewidth]{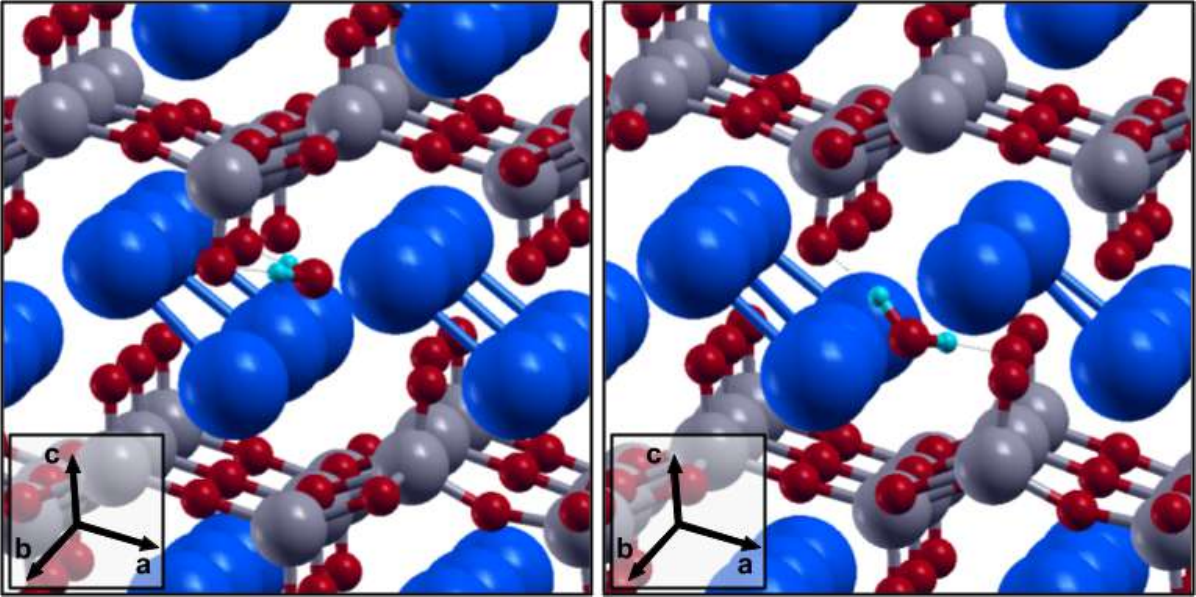}
\caption{\label{fig:bridging} XCrysDen \cite{Kokalj2003} illustration of the "RbO-coplanar" adsorption site (left) and the metastable "Bridging" adsorption site (right).}
\end{figure}

\begin{table}[h!]
\begin{center}
\begin{tabular}{|c||c|c|c|}
\hline
H$_2$O state & Gas phase & RbO co-planar & Bridging \\ \hline \hline
${\Delta E_{f}}$~(eV)  &   /  & -0.32 & -0.25 \\ \hline \hline
{d(O$^{(1)}$-H)}~({\AA}) &  / & 1.581 ($\times2$) &  1.632 / 1.573\\ \hline
{d(Rb-O$^{(w)}$)}~({\AA}) &  / & 2.725 / 2.700 ($\times2$) &  2.645 / 2.722 / 2.750  \\ \hline\hline
{d(O$^{(w)}$-H)}~({\AA}) &   0.967 ($\times2$) & 1.013 ($\times2$) &  1.005 / 1.014\\ \hline
 $\mathrm{H\hat{O}^{(w)}H}$ & 104.43 & 99.99 & 114.54\\\hline
\end{tabular}
\caption{\label{fig:tab_hyd_iso}\textbf{\underline{Computed properties of water molecule in gas phase and in adsorbed configurations:}} adsorption enthalpy $\Delta H_{ad}$; length of covalent bonds formed between the water molecule and the atoms of the structure; intra-molecular $\mathrm{H-O^{(w)}}$ bond lengths, internal angle $\mathrm{H\hat{O}^{(w)}H}$.}
\end{center}
\end{table}

\newpage
    
\begin{table}[h!]
    \centering
    \begin{tabular}{|c|c|c|c|}
    \hline
    \multicolumn{4}{|c|}{\textbf{RbO-coplanar configuration}} \\ \hline

        & \includegraphics[height=3.5cm]{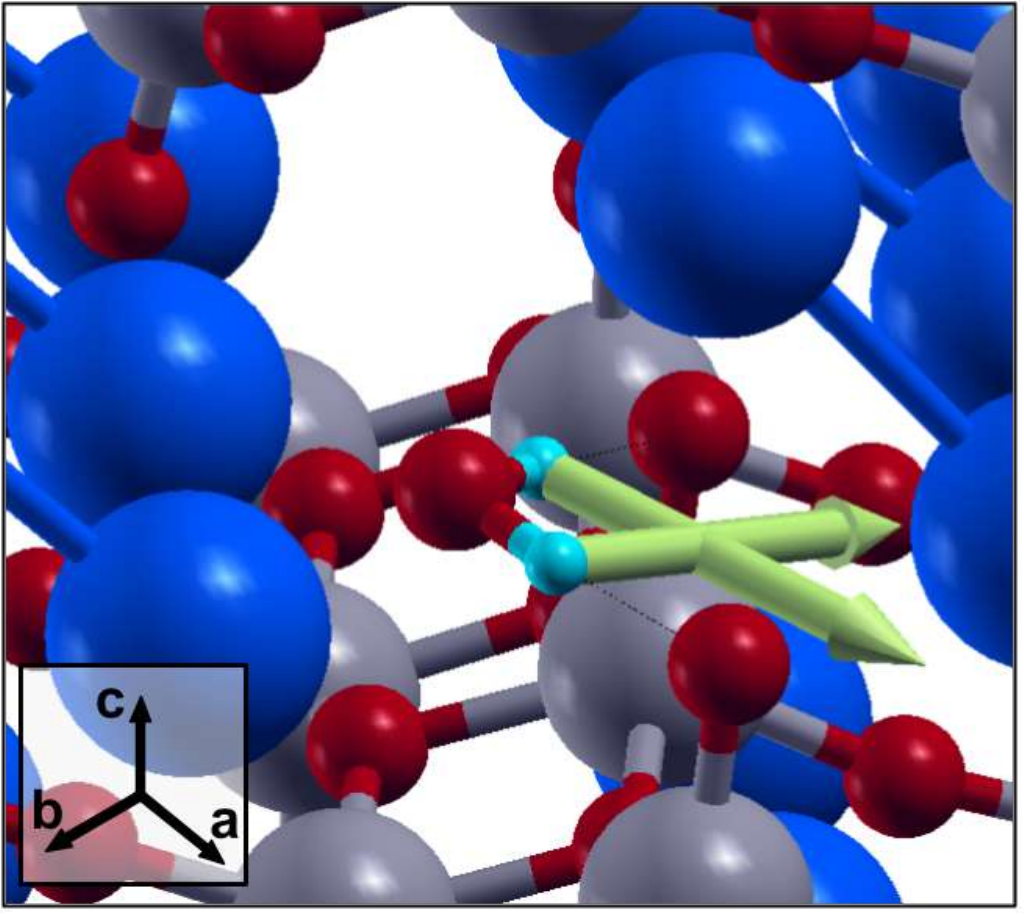} & \includegraphics[height=3.5cm]{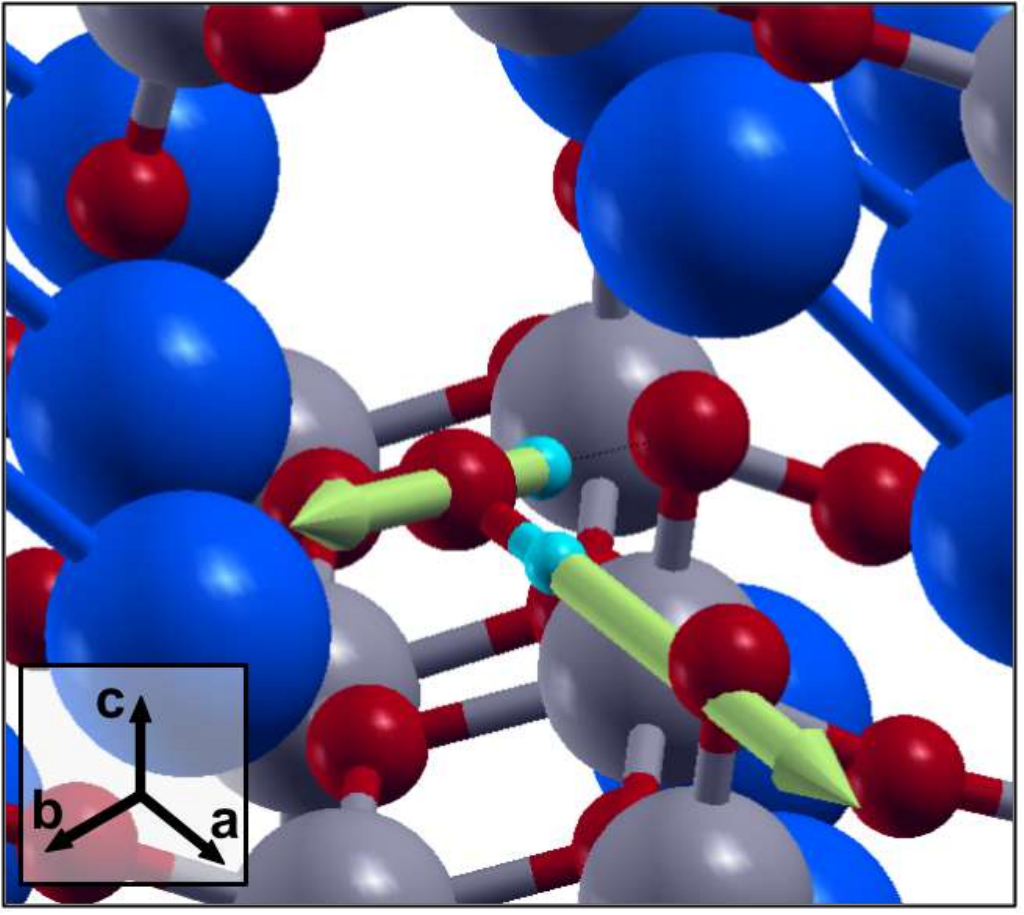} & \includegraphics[height=3.5cm]{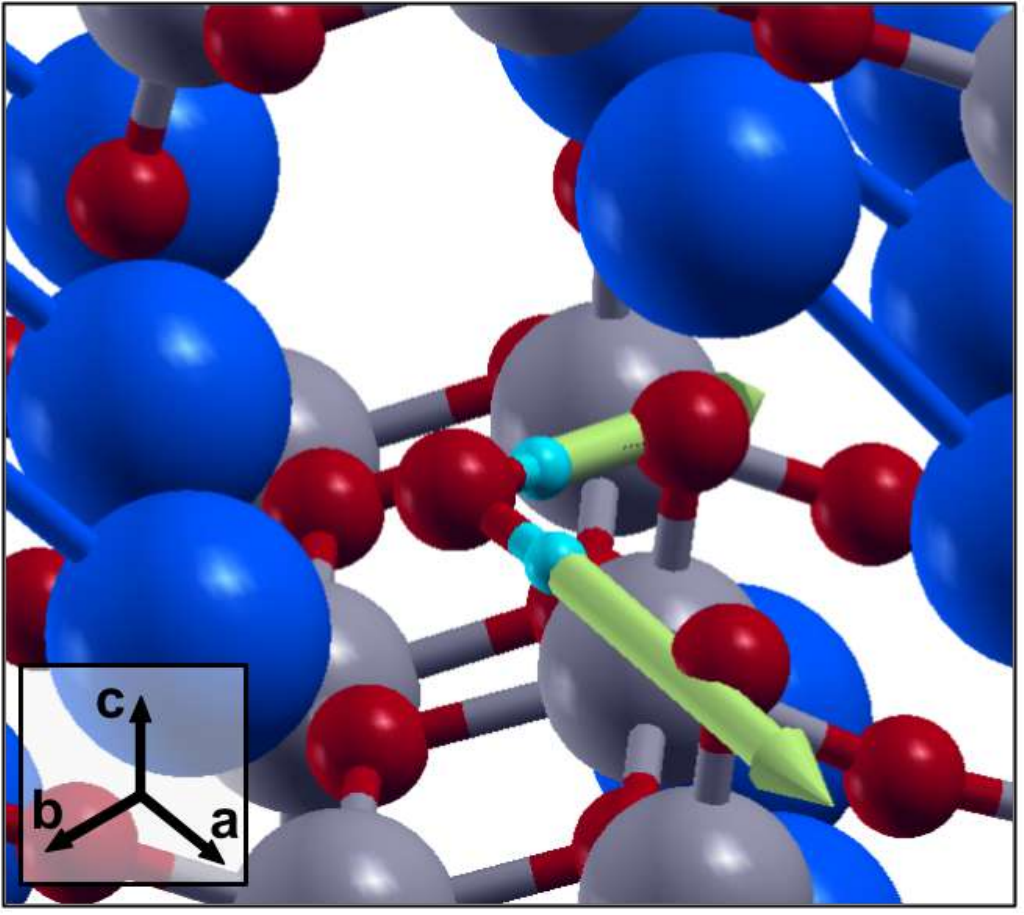} \\\hline
    Freq. (cm$^{-1}$) & 1658 & 2777 & 2879 \\\hline
     Nature    &  Bending & Antisym. Stretching & Sym. Stretching\\ \hline
     $\hat{\vec{p}}=(\hat{p}_a,\hat{p}_b,\hat{p}_c)$ & (-0.96,0.00,0.26) & (0.00,-1.00,0.00) & (-0.88,0.01,0.47) \\ \hline \hline
    \multicolumn{4}{|c|}{\textbf{Bridging configuration}}\\
    \hline
        & \includegraphics[height=3.5cm]{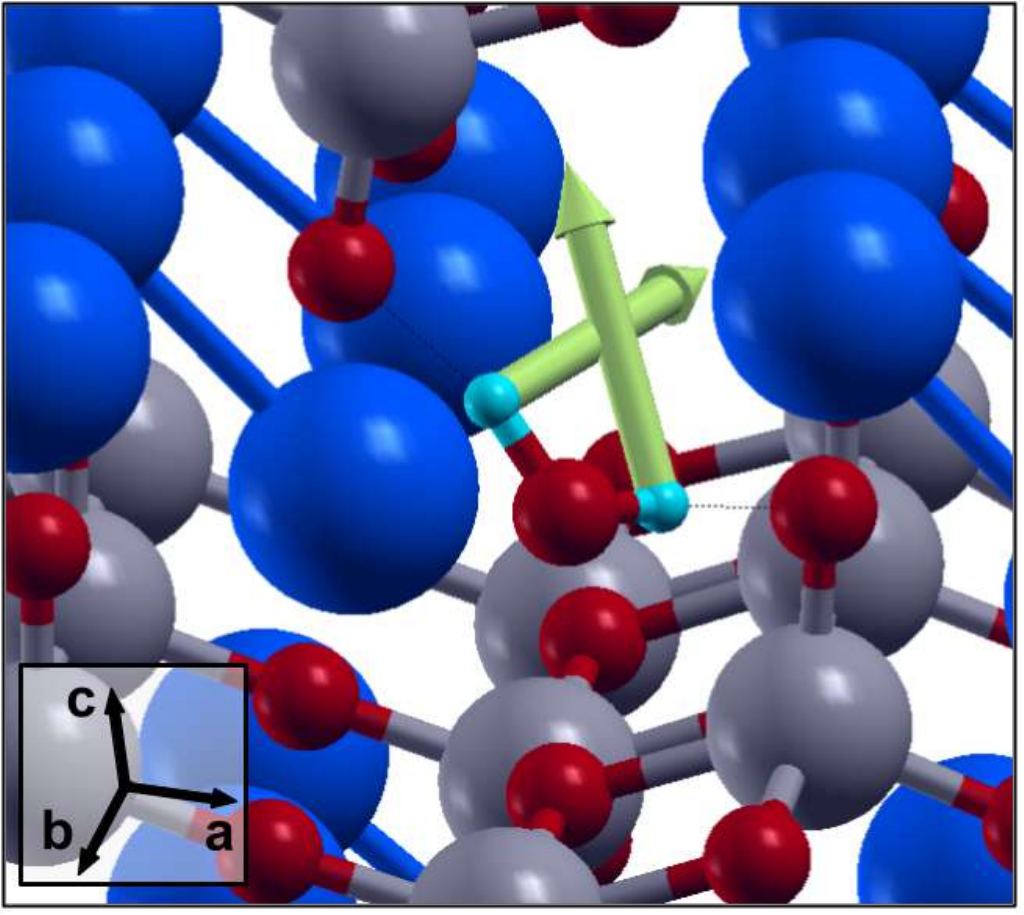} & \includegraphics[height=3.5cm]{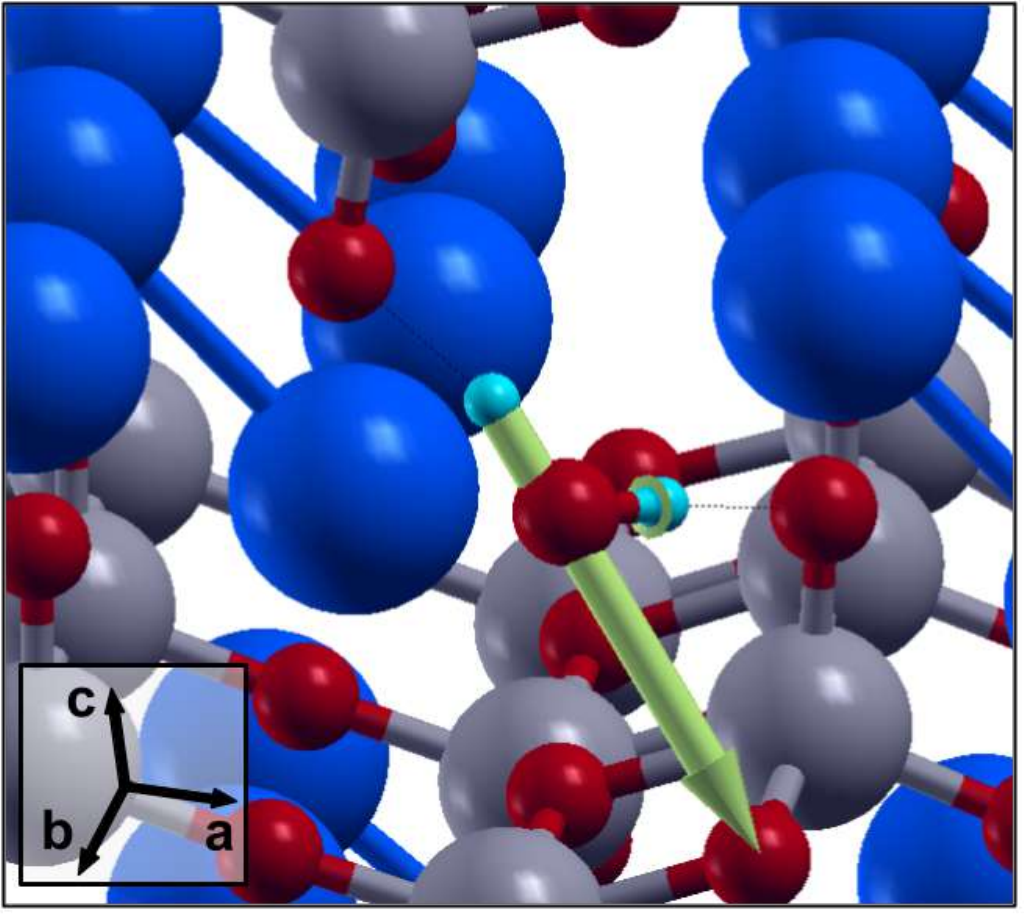} & \includegraphics[height=3.5cm]{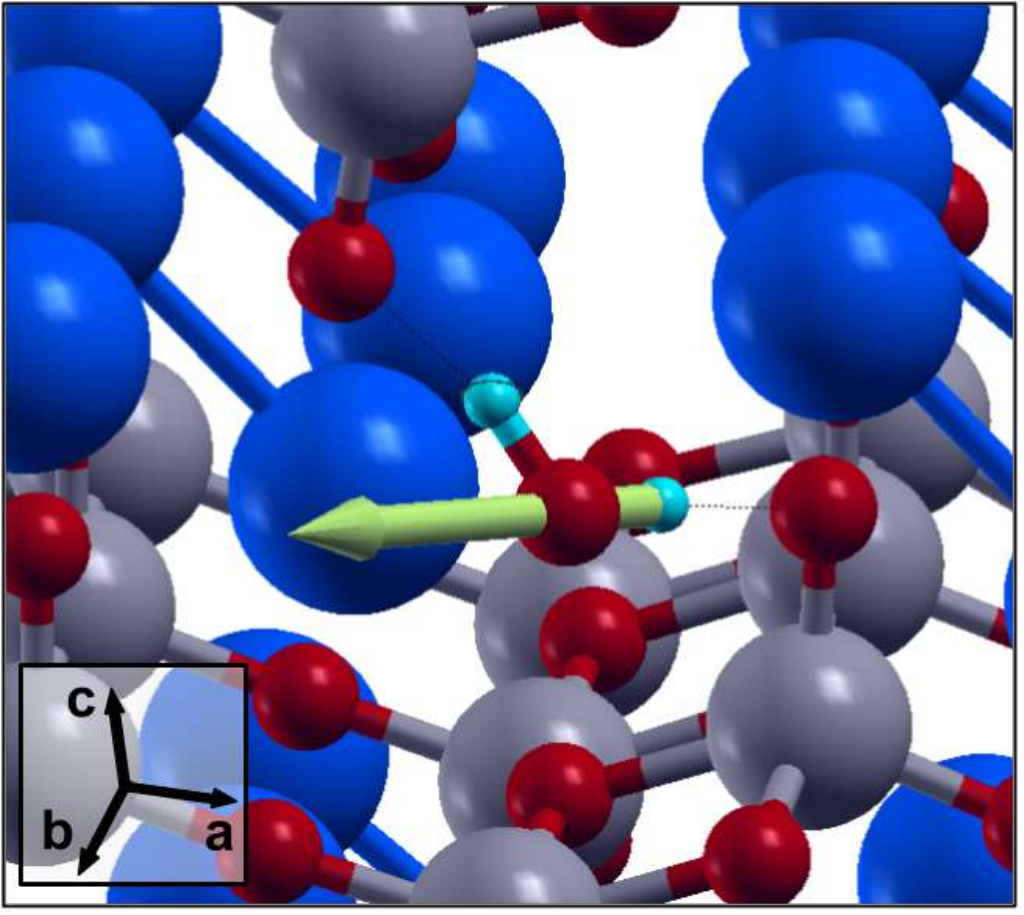} \\\hline
    Freq. (cm$^{-1}$) & 1505 & 2784 & 2925 \\\hline
     Nature    &  Bending & Stretching & Stretching\\ \hline
     $\hat{\vec{p}}=(p_a,p_b,p_c)$ & (0.47,0.83,0.29) & (0.46,0.06,-0.88) & (-0.64,-0.62,0.45) \\ \hline
     
    \end{tabular}
    \caption{\underline{\textbf{Vibrational modes of configurations with single water molecule adsorbed in the bulk.}} For each mode, XCrySDen \cite{xcrysden} visualizations illustrate the atomic displacement patterns, with yellow arrows indicating the directions of atomic motion. The corresponding vibrational frequencies (in cm$^{-1}$) and mode nature are reported, and the normalized polarization vectors are expressed in terms of the lattice vectors.}
    \label{tab:placeholder}
\end{table}

\newpage
\subsection{Organization of water molecules}
\begin{figure}[h!]
    \centering
    \includegraphics[width=0.7\linewidth]{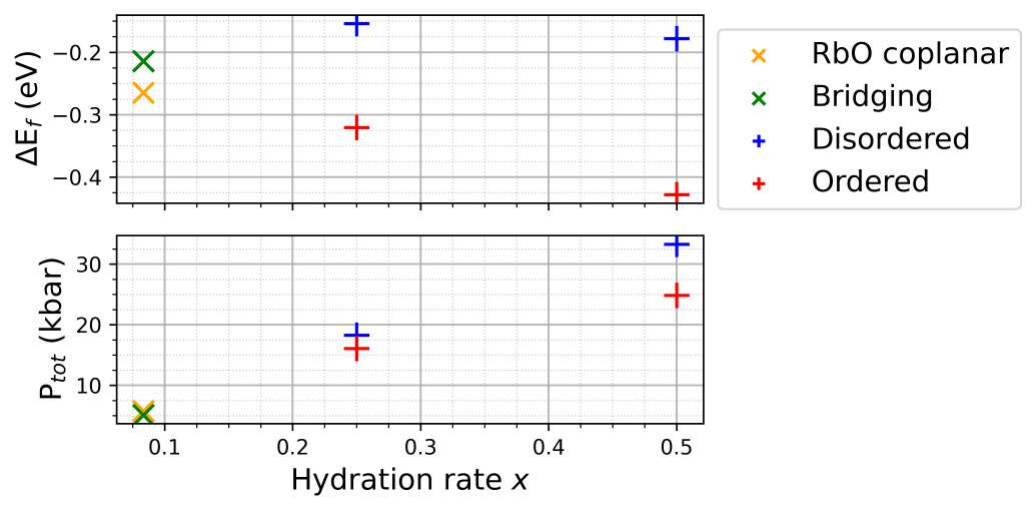}
    \caption{    \label{fig:ord_vs_disord}
     For hydrated systems $\mbox{Rb}_2\mbox{Ti}_2\mbox{O}_5:\left(\mbox{H}_2\mbox{O}\right)_x$, the formation energies in eV (upper panel) and the total constraint in kbar (lower panel) are evaluated at a volume V$_0$ corresponding to the non-hydrated volume. The lowest hydration level corresponds the isolated adsorbed molecules. At the intermediate hydration level (x=0.25), properties of water molecules arranged in an RbO-coplanar chain are compared to those of a randomly distributed configuration. At the highest hydration level shown (x=0.5), the arrangement of water molecules in two facing RbO-coplanar chains is compared to a randomly distributed configuration. It follows that ordered configurations are more stable than disordered ones, and also lead to slightly lower total stress.}
\end{figure}

\begin{figure}[h!]
\centering
\includegraphics[width=0.7\linewidth]{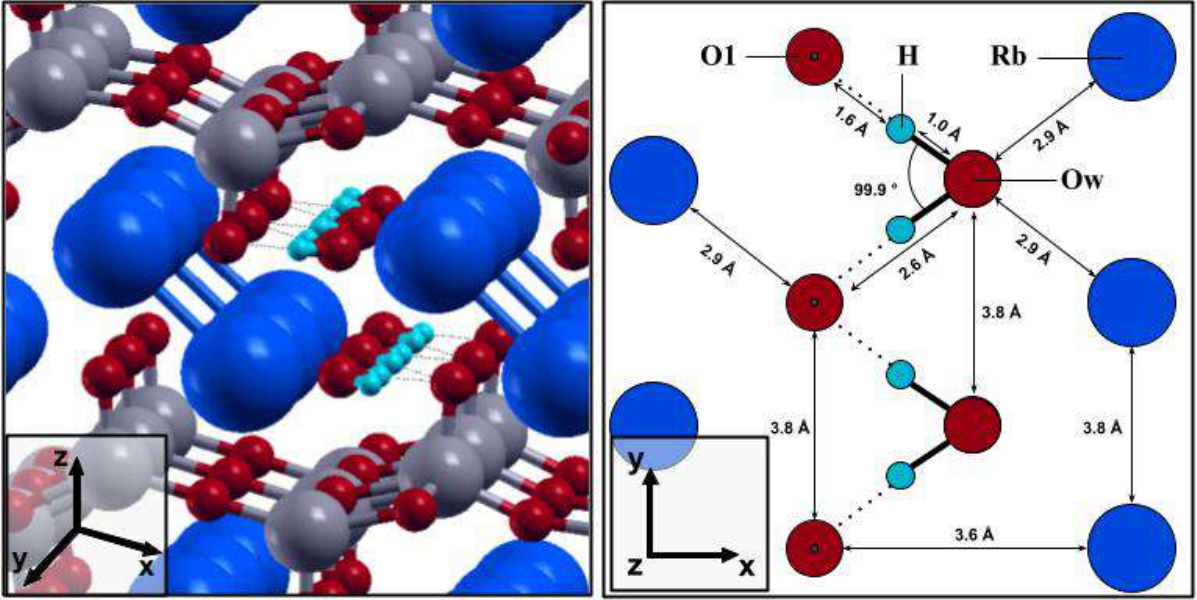}
\caption{\label{fig:double_chain_supp}\textbf{\underline{Illustration of double chains arrangement.}} The left figure illustrates the double chain arrangement described in the text using XCrysDen \cite{Kokalj2003}. The right figure is a schematic of a chain of water molecules viewed from the (001) plane, highlighting characteristic distances.}
\end{figure}

\begin{figure}[h!]
    \centering
    \includegraphics[width=0.9\linewidth]{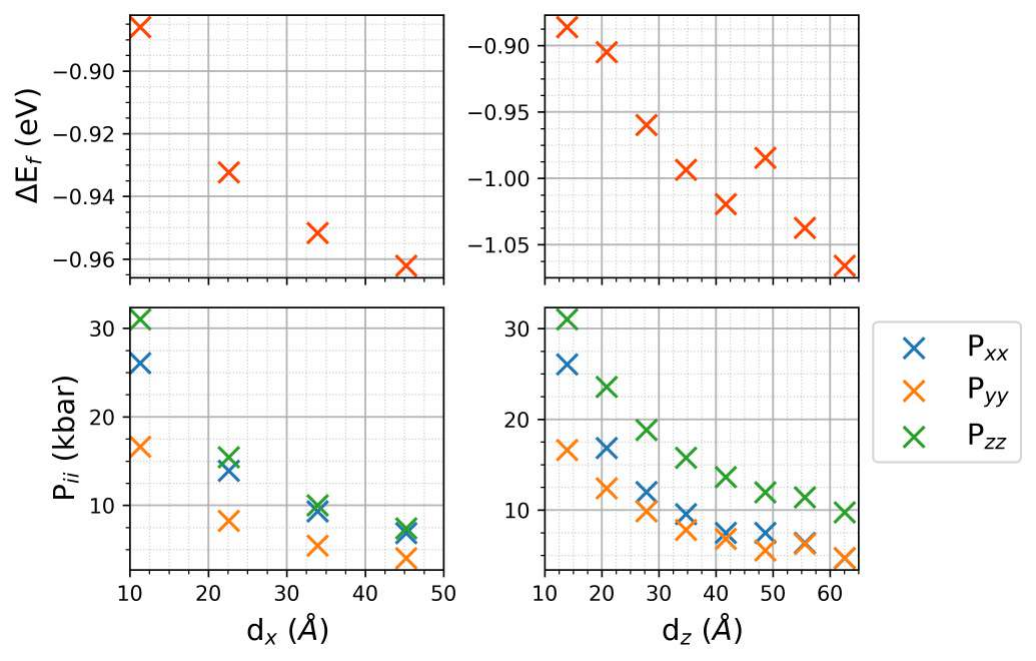}
    \caption{\label{fig:chains_distrib}\textbf{\underline{Hydrated systems with double chains distributed at different distances in a given direction:}} For super-cells N$_x\cdot$N$_y\cdot$N$_z$ featuring a double chains arrangement, N$_y$ is fixed to 1, the left panels correspond to systems with increasing N$_x$ and N$_z$=2 while the right panels correspond to systems with N$_x$=1 and increasing N$_z$. In each case, the energy of formation (top panels) is given for a system with a fixed volume along with the main components of the internal stress (lower panels). It reveals an effective repulsive interaction in both $x$ and $z$ directions, suggesting a long-range distribution of these double chains arrangements.}
\end{figure}

\begin{figure}[h!]
    \centering
    \includegraphics[width=0.6\linewidth]{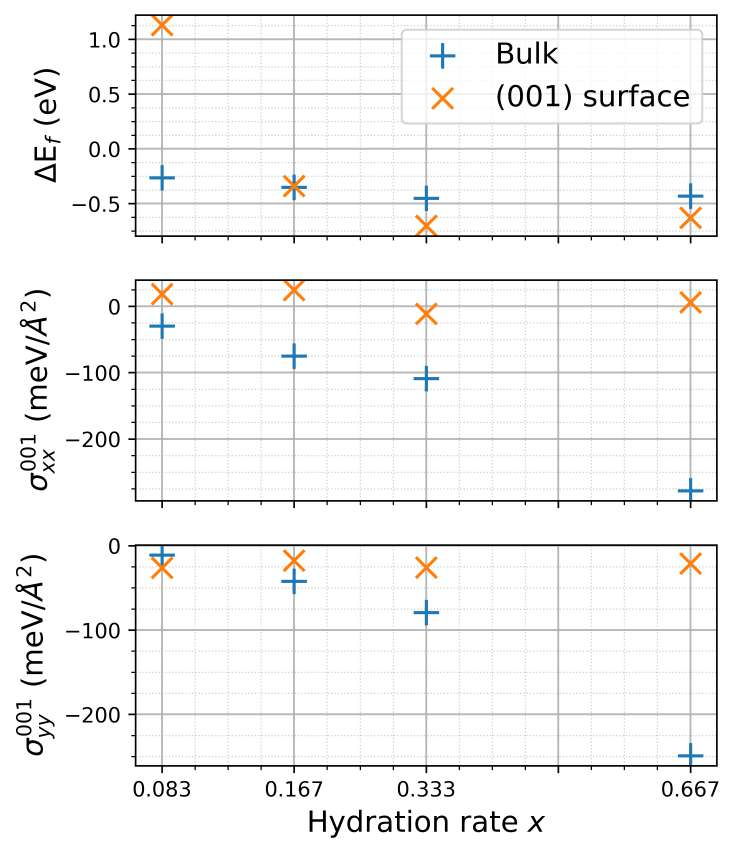}
    \caption{\label{fig:bulk_vs_surf}Comparison of the properties of systems hydrated in bulk with a constant volume (blue) and at the (001) surface (orange). 
    For different hydration rates the formation energy (in eV) is represented in the top panel, the stresses on the (001) plane in the x and y directions (in meV/$\r{A}^2$) are respectively represented in the middle and lower panels.}
\end{figure}

\clearpage
\section{Ionic conductivity measurements}


\begin{figure}[h!]
\begin{center}
\includegraphics[width=0.45\linewidth]{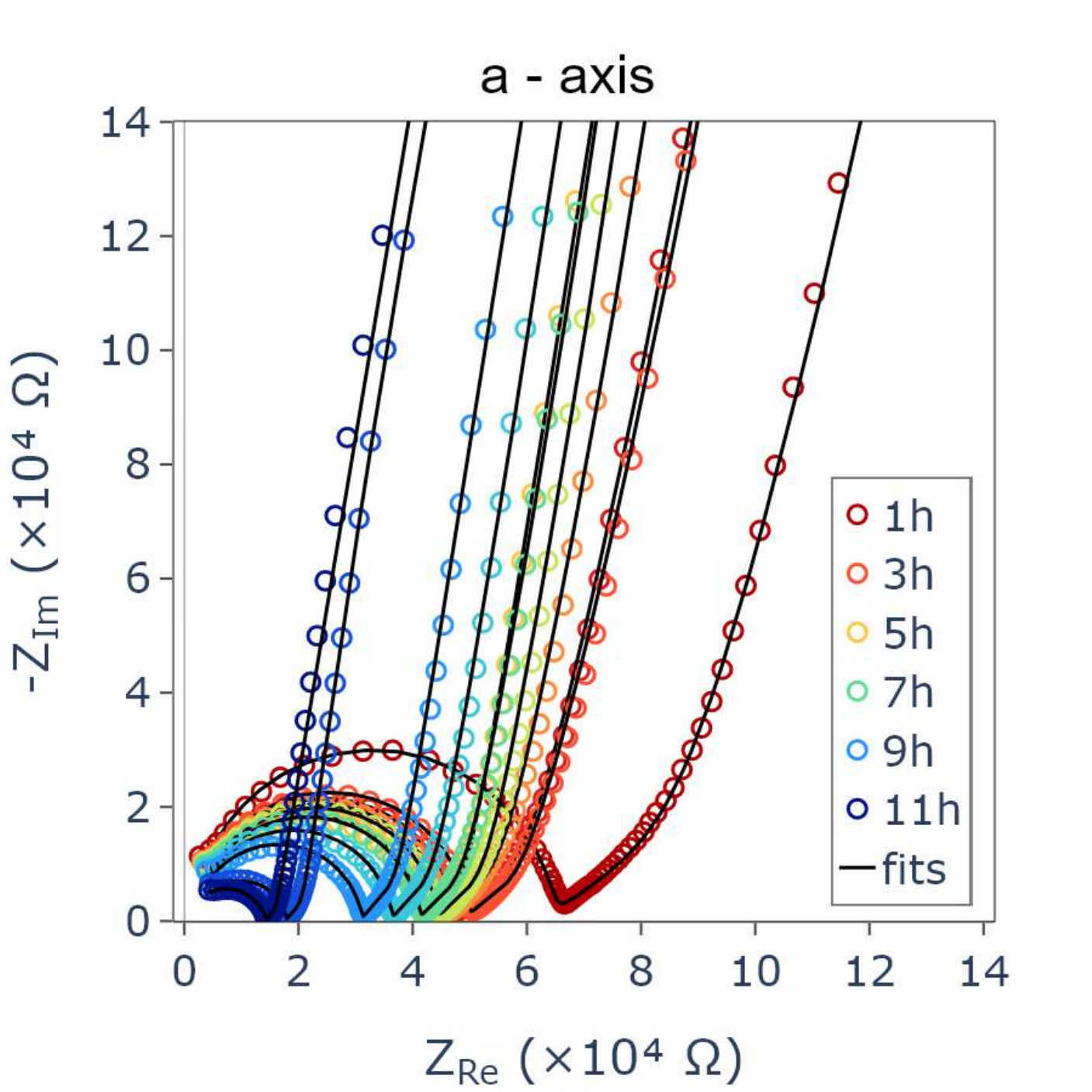}
\includegraphics[width=0.45\linewidth]{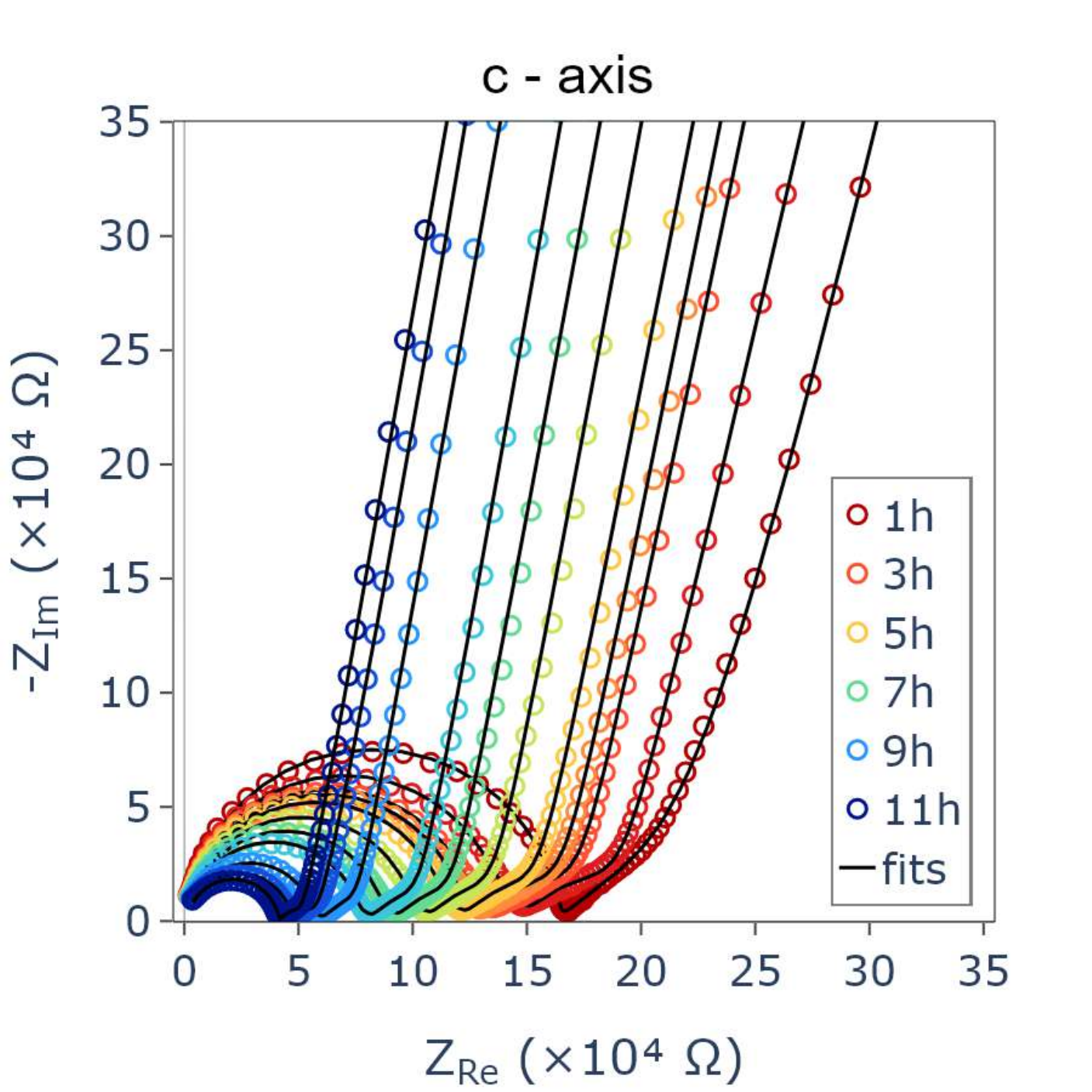}
\caption{\label{Nyquist_ac} Nyquist diagrams of the impedance measured along $\vec{a}$ for the left panel and along $\vec{c}$ for the right panel for a pristine RTO sample as function of time exposure to the laboratory atmosphere. The black lines indicate the fits obtained from the equivalent electrical circuit discussed in the text (Figure \ref{Nyquist_circuit})}
\end{center}
\end{figure}

\begin{figure}[h!]
\begin{center}
\includegraphics[width=0.45\linewidth]{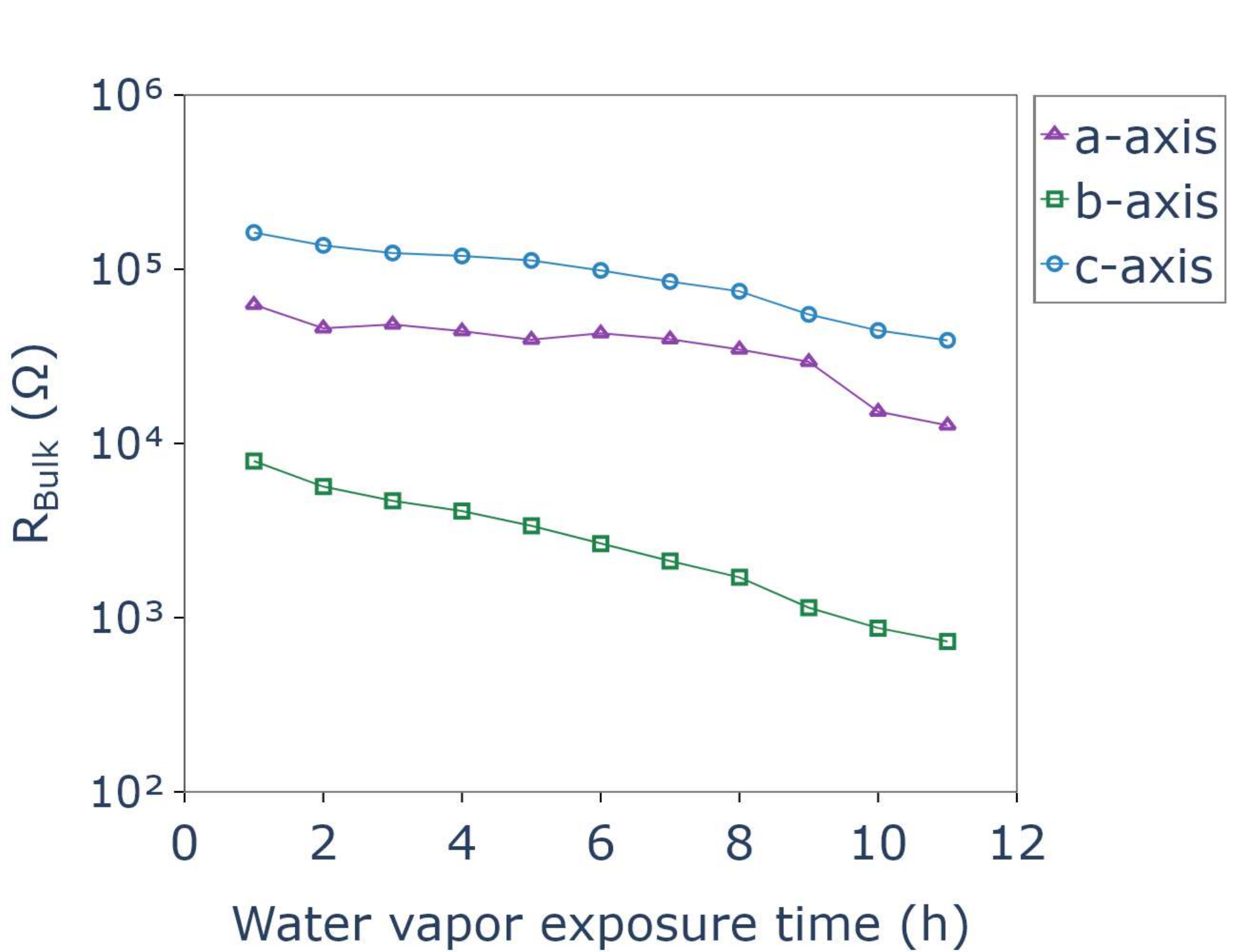}
\includegraphics[width=0.45\linewidth]{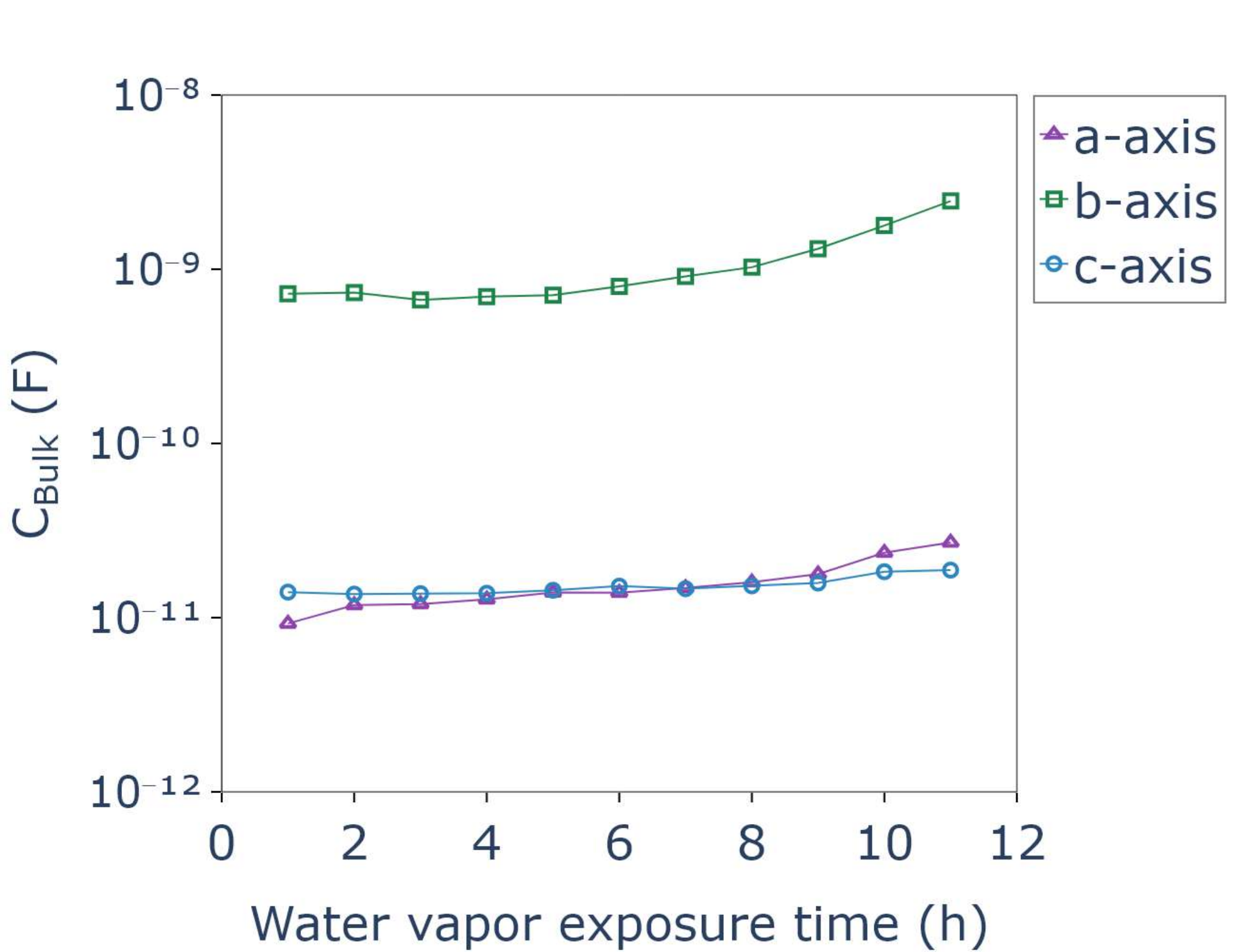}
\caption{\label{Nyquist_R_C} Right : R$_{Bulk}$ as a function of exposure time to the laboratory atmosphere for the three crystallographic orientations. Left : C$_{Bulk}$ as a function of exposure time to the laboratory atmosphere for the three crystallographic orientations. R$_{Bulk}$ and C$_{Bulk}$ are collected from the fit of the Nyquist diagrams from the equivalent circuits detailed in the text. }
\end{center}
\end{figure}



\end{document}